\documentstyle[prl,aps,preprint,tighten,floats,aps,epsfig,psfig]{revtex}
\def\simlt{\stackrel{<}{{}_\sim}}
\def\simgt{\stackrel{>}{{}_\sim}}

\def\btabl{\begin{table}}   \def\etabl{\end{table}}
\def\bea{\begin{eqnarray}}   \def\eea{\end{eqnarray}}
\def\bnn{\begin{eqnarray*}}   \def\enn{\end{eqnarray*}}
\def\beq{\begin{equation}}   \def\eeq{\end{equation}}  
\def\btabu{\begin{tabular}}   \def\etabu{\end{tabular}}
\def\bec{\begin{displaymath}} \def\eec{\end{displaymath}}
\def\nn{\nonumber}
\def\eqref#1{(\ref{#1})}
\renewcommand{\baselinestretch}{1.2}

\begin{document}
\draft
\date{\today}
\preprint{\vbox{\baselineskip=13pt
\rightline{CERN-TH/98-40}
\rightline{FAMNSE-97/20 
}
\rightline{LPTHE Orsay-97/69}
\rightline{hep-ph/9802393}}}
\title{Finite-size effects on multibody neutrino exchange}
\author{As. Abada$^{a}$, 
O. P\`ene$^b$ and J. Rodr\'\i guez-Quintero$^c$ \footnote{
e-mail: abada@mail.cern.ch, jquinter@cica.es, 
pene@qcd.th.u-psud.fr.}} 
\vskip -0.5cm
\vspace{-0.5cm}
\address{{\small 
$^a$ Theory Division, CERN, CH-1211 Geneva 23, Switzerland.\\
$^b$ Laboratoire de Physique Th\'eorique et Hautes
Energies\\
Universit\'e de Paris XI, B\^atiment 211, 91405 Orsay Cedex,
France.
\\
$^c$ Departamento de F\' \i sica At\'omica, Molecular y Nuclear, Universidad 
de Sevilla \\ P.O. Box 1065, 41080 Sevilla, Spain.}}

\vskip -1cm

\maketitle \begin{abstract}  

The effect of 
 multibody massless neutrino exchanges
 between neutrons inside a finite-size
neutron star  is studied. We use an effective Lagrangian,  
which incorporates the effect of the neutrons on the neutrinos.
Following Schwinger, it is shown that the total interaction energy
density is computed by comparing the zero point energy of the neutrino sea
with and without the star. It has already been shown that in an infinite-size 
star the total energy due to neutrino exchange vanishes exactly. The opposite
claim that massless neutrino exchange would produce a huge energy
is due to an improper summation of an infrared-divergent quantity. 
The same vanishing of the total energy has been proved
 exactly in the case of a finite star in a one-dimensional toy model.
 Here we study  the three-dimensional case. 
 We first consider the effect of a sharp
 star border, assumed to be a plane. We find that there is a non-vanishing 
 of the zero point energy density difference between the inside and the outside due to 
 the refraction index at the border and the consequent non-penetrating 
  waves.  
  An analytical and numerical calculation for the case of a spherical 
 star with a sharp border confirms that the preceding border effect is the
 dominant one. The total result is shown to be infrared-safe, thus confirming 
 that there is no need to assume a neutrino mass. 
 The ultraviolet cut-offs, which correspond in some sense to the matching of
 the effective theory with the exact one,  
  are discussed. Finally the energy  due to long distance neutrino 
 exchange is of the order of 
 $10^{-8}$--$10^{-13}\ 
 {\mathrm{GeV}\hbox{ per neutron}}$, i.e.  
negligible with respect to the neutron mass density. 

\end{abstract}

\leftline{} 
\leftline{CERN-TH/98-40}
\leftline{February 1998}  

\renewcommand{\baselinestretch}{1.2}

\newpage \section{Introduction}
The  massless neutrino exchange interaction between neutrons, protons, etc.,  
 is a long-range force \cite{Fein68}--\cite{sikivi}.
In a previous work \cite{rescue}, the long-range interaction effects
 on the stability of a neutron star
due to multibody exchange of massless neutrinos have been studied. 
We  have
shown that the total effect of the many-body forces of this type results in an
infrared well-behaved contribution to the energy density of the 
star and that it is negligible with respect  to the star mass density. This is 
  in agreement with two recent non-perturbative calculations done by
 Kachelriess \cite{Kachelriess} and by Kiers and  Tytgat \cite{kiers}.

 This work
 is in contradiction with the repeated claim by Fischbach \cite{fischbach}
  that, unless the 
 neutrino is massive, neutrino exchange renders a neutron star unstable, 
 as the induced self-energy exceeds the mass of the star because of the 
 infrared effects associated to neutrino exchange between four or more neutrons.
  In our opinion, the latter ``catastrophic'' result is a consequence
  of summing up large infrared terms in perturbation outside the radius of
  convergence of the perturbative series. The non-perturbative
    use of an effective Lagrangian
  immediately gives the result without recourse to the perturbative series, and
  the result is small. 
  
Smirnov and Vissani \cite{smirnov}, following the same method as  in 
ref. \cite{fischbach}, summing up multibody exchange contributions order by
order, showed that the 2-body contribution is damped by the blocking effect of
the neutrino sea \cite{loeb}. They guessed that this damping would apply to
many-body contributions, and hence would reduce the catastrophic effect claimed 
by Fischbach.
In our previous work \cite{rescue}, we also considered  the effect of the neutrino 
sea  
inside the neutron star. This effect has been introduced in our 
non-perturbative
calculation by using Feynman propagators of neutrinos inside a dense medium, 
which incorporate the condensate term. We noticed that this condensate
is present, but in our opinion it is not the most important of the effects  that neutralize
 the
catastrophic effect expected by Fischbach, since it only brings a tiny change
 to the non-perturbatively summed interaction energy, leaving unchanged our
  conclusion that the weak self-energy is infrared-safe.

 We have also stressed in \cite{rescue} that the neutrino condensate was 
 related to the existence of a
  border
  \footnote{
  At that point, we should call  the reader's attention to  a minor mistake that 
  was made 
  in the previous  calculation \cite{rescue}: a pole was forgotten in the 
  calculation of
  the weak self-energy, and its contribution is exactly annihilated
   by 
  the condensate's, as shown in ref. \cite{note}: the computation of 
  the self-energy gives zero when the forgotten pole is  taken 
  into account.}.
   This was  demonstrated 
 in the ($1\  +\  1$)-dimensional star in \cite{note}: the blocking
  effect, which implies 
 the trapping of the neutrinos inside the star  
  while the antineutrinos are repelled from it, 
  is a natural consequence of 
  the existence of a border. 
  Indeed a proper treatment 
  of the effect of the border automatically
 incorporates  the condensate contribution as a
  consequence of the appropriate boundary conditions for the neutrino 
  Feynman propagator
  inside the star.

Since our treatment directly incorporates the neutrino sea effect and,
 on the other
hand, since we stick to our strategy of directly computing
 the total neutrino
interaction energy by using an effective Lagrangian, we have in a sense thus 
generalized 
 the result of Smirnov and Vissani \cite{smirnov} because our result holds in a
 non-perturbative way and accounts for all the $n$-body
 contributions, while theirs holds for $2$-body contributions only.
  
  It has been  objected \cite{smirnov1} to our study \cite{rescue}  that 
 we  worked in an
 approximation where we  neglected the border of the neutron star.
  Our belief is that this simplifying hypothesis does not change the
 fundamental result that the total effect  of the  multiple neutrino
  exchange  to the 
 energy density of the star is not infrared-divergent.

Indeed, in \cite{note} we have proved that, in $(1\  +\  1)$
 dimensions {\it with borders}, the result of \cite{rescue} for the 
 infinite star {\it without borders}
 is kept unchanged: the net interaction energy due to
  long-range neutrino exchange is exactly zero. This result was obtained by
  computing Feynman vacuum loops with neutrino propagators derived from
   the effective Lagrangian, which incorporates the neutron interaction. The
   latter propagators are not translational-invariant, because of the star 
   borders.
   We have found a physically simple explanation for the vanishing of the net
   interaction energy. It relies on the fact that the negative energy states, 
   in
   the presence of the star and without the star, are in a one-to-one
   correspondence and have exactly the same energy density. It results that 
   the zero point energy is the same with and without the star, 
   for any density profile of the star.

The main goal of 
 the present work is to follow on 
 taking into account the finite-size and border effects.
 Two main conclusions of ref. \cite{note} 
  are useful for 
the ($3\ +\ 1$)-dimensional star:
 i) the natural connection between the neutrino sea and the border, 
  ii) the correct definition of the
zero-energy level of the Dirac sea.

 From \cite{note}, we know that the zero-energy level of the Dirac sea has to be
adjusted by comparing the  asymptotic behaviour of the wave functions far
outside the star with the free solutions in the absence of the star. From there, we
know the correct $i \epsilon$ prescription, which has to be imposed in the
propagators of the neutrinos in the presence of the star; we could in
principle compute the closed loops to get the vacuum energy density. 
However, this approach is technically very difficult. A simpler method, 
the derivation of which is  recalled in section \ref{general}, is 
to simply add the energy density of the negative energy solutions
in the presence of the star, minus the same in the absence of the star.

The vanishing of the  neutrino exchange energy density in the star, found in the
case of an infinite star \cite{rescue} and in $(1\ +\ 1)$ dimensions \cite{note},
which will be summarized in section \ref{1+1},   
 is not valid in $(3\ +\ 1)$ dimensions. The main reason for that will
    be illustrated in section \ref{flat}  by zooming to the border effect, i.e. 
    considering a
   plane border in $(3\ + \ 1)$ dimensions. There is a non-trivial refraction index
   that modifies the wave energy densities as they penetrate the star. Some waves are
   forbidden to penetrate  and this yelds the dominant
   contribution.

    In section \ref{spherical}, we perform  analytical and numerical
calculations, which take into account
  the curvature of the border and use a
 spherical star with a sharp border.  We find a very simple approximate 
 formula for the zero point neutrino energy density in the star, 
 and demonstrate numerically the validity of this approximation. 
 It results  that indeed the neutrino-induced energy density in the star
 does not vanish and is dominantly explained  by the above-mentioned border 
 effect.

In any case, these non-vanishing neutrino exchange energy densities 
are all perfectly regular in the
infrared and do not present any resemblance to Fischbach's effect.
 On the other
hand they are ultraviolet-singular. This is not unexpected  since anyhow our
effective Lagrangian is only valid below some energy scale where the neutrons
may be considered at rest. In section \ref{discussion}, we also discuss in  
some detail the effect of
decoherence for distances larger than the neutrino mean-free path, which
smoothes down the ultraviolet singularity.
  
 We  conclude   that the stability of compact and dense
  objects such as a white dwarfs, neutron stars, etc.,  
   are not affected by the neutrino exchange, 
 even if  neutrinos are massless.

\section{General Formalism}\label{general}
In our calculations, we assume  that the material of the  neutron star is made
exclusively of neutrons, among which neutrinos are exchanged.
The density-dependent corrections to the neutrino self-energy result, at
leading order, from the evaluation of
$Z^0$-exchange diagrams between the neutrino and the neutrons in the
medium, with the $Z^0$ propagator evaluated at zero momentum. 
The vacuum energy--momentum relation for massless
fermions, $E=|\vec q|$, where $E$ is the energy and $|\vec q|$ the
 magnitude of the momentum vector, does not hold in a medium \cite{wolf}. In
 our case, 
following refs.\cite{nieves} and \cite{note}, they
can be summarized by the following dispersion 
relations:

\bea \begin{array}{ll}
E_{\nu} = q_0 =|\vec q| +b \ \ \ \quad\quad \ \ \ \ \mbox{for}\  |\vec q| > |b|\\
E_{\overline{\nu}}=-q_0= \left\{ \begin{array}{ll}
|\vec q| - b  \ \  & \quad \forall |\vec q|\ \ \  \mbox{or}\ ,  \\
-|\vec q| - b \  &\ \ \  \mbox{for} \ |\vec q| < |b|
\end{array} \right.\end{array} 
\
\label{disp}
\eea
where
\bea
b\simeq - \sqrt{2} G_F\, n_n/2 \sim -0.2\, 10^{-7}\,{\rm GeV}\sim -10^{-7}
\,{\rm fm}^{-1},\ {\mathrm{and}}\ n_n\sim 0.4 \ {\rm fm}^{-3}.
\label{pmb}
\eea

In this paper we will use a  star radius of 

\bea
 R\simeq 10 \ {\mathrm{km}}, 
\qquad {\mathrm{whence}}\qquad bR \simeq - 10^{12}\ \ .
 \label{huge}\eea
 In eq. (\ref{disp}), $b$ summarizes the zero-momentum transfer interaction 
 of a massless neutrino
with any number of neutrons present in the media. Sensibly enough, it
depends on the neutron density  which will be assumed to be constant for
simplicity. When a neutrino sea is present \cite{loeb}
  \footnote{
    $\ b$ is given by:
\bea
  b\simeq - \sqrt{2} G_F\, (n_n-n_{\nu})/2
- \frac{8\sqrt{2} G_F \kappa}{3m_Z^2}n_{\nu}\langle E_\nu\rangle\ ,\nn
\eea
where $\langle E_\nu\rangle$ is the  average neutrino energy of the  
medium, defined in its 
rest frame, and $n_\nu$ the neutrino density ($n_\nu\sim 4\ \times 
10^{-23} \ \mathrm{fm}^{-3}$). These corrections to the fermion propagation 
would
give higher-order effects, and we disregard them in the present work.}, 
  the neutrino condensate does not sensibly modify the 
value of $b$ \cite{Kachelriess}.

In the approximation where the neutrons are static, in the sense that they do 
not feel
 the recoil from the scattering of the neutrinos, we can study the neutrino exchange 
 through the
 following effective Lagrangian as done in ref. \cite{rescue}:
 \bea
 {\cal {L}}_{\mathrm{eff}}=i{\overline{\nu}} \partial\!\!\!/ \ \nu ( r) 
 -b\ { \overline {\nu}}\gamma_0\nu \ \theta(R -r )\ ,\label{lageff}\eea
where $R$ is the radius of the neutron star.

 The dispersion relations in eq. (\ref{disp}) show a displacement of the energy levels for the
different modes, a negative shift for neutrinos, and a positive one for
antineutrinos; the Dirac sea level is displaced. Would the neutron star
occupy the whole Universe, it
would just mean a change of variables, with no physical consequence.
The finite size of the star changes the picture.
 Notice that $b$ acts as the depth of a potential well, as it will be
 considered in section \ref{spherical}. 
 It is repulsive for antineutrinos
and attractive for neutrinos, which then condense.

A traditional way to estimate the energy induced by neutrino exchange is to
compute first the  exchange potentials involving  $2,\ 3, \ 4, ...$  neutrons  
and then add their
contributions integrated over the neutron positions in the star.
 This is
the way chosen by Fishbach { et al.} \cite{fischbach} to compute the weak 
self-energy.
The drawbacks of this method are the following:\\
i) The calculation turns out to be so difficult that many approximations are 
necessary. \\ii) It assumes implicitly that every neutron may interact only 
once with neutrinos. iii)
 The resulting
interaction energy increases with the number $n$ of neutrons involved grossly 
as
 $(bR)^n\simeq 10^{12 n}$ and eventually 
becomes very large.
In fact {\it it would go to infinity} if these authors did not stop when $n$ 
equals the number of neutrons of the star, as a consequence of the 
above-mentioned unjustified assumption that a given neutron cannot interact
 more than once.   The large
 parameter $bR$ (eq. (\ref{huge})) reflects the infrared-sensible behaviour
  of each term in the
 series.
The main problem of this approach is that the summation is done outside  
the radius of convergence of the 
	  perturbative series, 
	  leading to an unacceptably huge result \cite{fischbach}. 
	  In ref. \cite{rescue}, 
	  we used 
 instead  a simpler and more direct method, which is non-perturbative 
 and based on an effective action. This method does not involve
  uncontrollable approximations, it does incorporate automatically 
  multiple interactions of one same neutron and finally it leads to 
  a totally reasonable result. 
 This method is also followed in recent works, refs. \cite{Kachelriess}
  and  \cite{kiers}. 
 We should insist that, in this approach, there is no extra assumption 
 added to 
 the problem.

In our work, we use the 
Schwinger tools \cite{schwinger} in order to
 compute the density of weak interaction energy $w(\vec x)$ due to the multibody neutrino
 exchange. It is
 given by  the  difference between the energy density for a neutrino
 propagating in the ``vacuum'' defined by the neutron star,
 $|{\hat{0}}\rangle$, and the corresponding one for the real, matter-free, vacuum,
 $|0\rangle$:

\bea
 w(\vec x)\equiv\langle\hat{0}|{\cal H}(\vec x)|{\hat{0}}\rangle - 
 \langle{0}|{\cal H}_0(\vec x)|{0}\rangle\ ,
\label{energy}
\eea
where ${\cal H}(\vec x)$ and ${\cal H}_0(\vec x)$ are the Hamiltonian densities of 
free
and interacting neutrinos. 
Concretely, to compute  this
 weak interaction energy analytically,  using \cite{schwinger}, we can write it as: 
 \bea
 w(\vec x)= -i{\partial\over\partial x^0 }
 {\mathrm{tr}}\left\{\gamma_0
 \left [ S_F( x,  y)-S_F^{(0)}(x, y)\right ]\right \}_{ y\to x}.
 \label{Wanalyt}
\eea
The r.h.s. does not depend on  $x^0$ from time-translational invariance. 

\begin{figure}[h]   
\begin{center}
$
\epsfbox{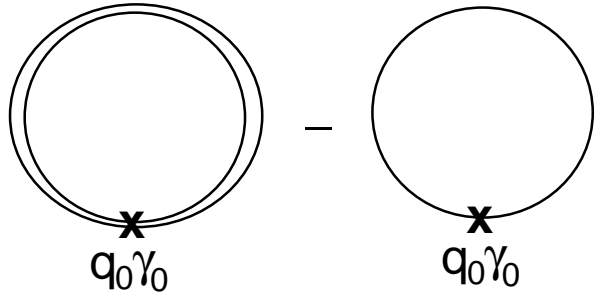}
$
\caption[]{\it{Schematic representation of $w$, eq. (\ref{Wanalyt}).
The simple and double line represent the neutrino propagator, 
in the matter-free vacuum and in the vacuum of the interacting 
medium, respectively.}} \protect\label{diagramatic}
\end{center}
\end{figure}

In diagrammatic formulation, the formal  eq.  (\ref{Wanalyt})  corresponds
 to the
computation of the difference of the diagrams in fig. \ref{diagramatic}.

\noindent Using the notation 
$$\sum\!\!\!\!\!\!\!\int _{n_\pm}\ \equiv\ \sum_{j_z,l,...}\int dE_{n_\pm}\ 
$$ 
and from the definition
\bea
 S_F(x, y)\gamma_0&=&\theta(x^0-y^0)\sum\!\!\!\!\!\!\!\int _{n_+}
 \Psi_{n_+}(\vec x)\Psi^{\dag}_{n_+}(\vec y)\,e^{-iE_{n_+}(x^0-y^0)}\cr
&-&\theta(y^0-x^0)\sum\!\!\!\!\!\!\!\int _{n_-}
 \Psi_{n_-}(\vec x)\Psi^{\dag}_{n_-}(\vec y)\,e^{-iE_{n_-}(x^0-y^0)}\ ,
\eea
where $E_{n_-}$ ($E_{n_+}$)
 is the negative (positive) energy of the eigenstate: 
$H \Psi_{n_\pm} = E_{n_\pm} \Psi_{n_\pm}$, 
  $w(\vec x)$ can be rewritten \cite{schwinger} as:
\bea
w(\vec x)=\sum\!\!\!\!\!\!\!\int _{n_-} E_{n_-} \Psi^{\dag}_{n_-}(\vec x)
\Psi_{n_-}(\vec x)
-\sum\!\!\!\!\!\!\!\int _{n_-}E^{(0)}_{n_-} 
\Psi^{\dag(0)}_{n_-}(\vec x) \Psi^{(0)}
_{n_-}(\vec x) \ ,\label{somme}
\eea
\noindent where $(0)$ refers to the matter-free vacuum and 
where we have taken
the limit  $y^0 \to x^0$ with $y^0 > x^0$. 
Taking the average of the limits $y^0 \to x^0$ with $y^0 > x^0$
and $y^0 \to x^0$ with $y^0 < x^0$, as done in \cite{kiers},  would
lead to:
\bea
 w_{sym}(\vec x)&=&\frac 1 2 \sum\!\!\!\!\!\!\!\int _{n_-} E_{n_-} 
 \Psi^{\dag}_{n_-}(\vec x)
\Psi_{n_-}(\vec x)
-\frac 1 2\sum\!\!\!\!\!\!\!\int _{n_-}E^{(0)}_{n_-} 
\Psi^{\dag(0)}_{n_-}(\vec x) \Psi^{(0)}
_{n_-}(\vec x) \cr
&-&\frac 1 2 \sum\!\!\!\!\!\!\!\int _{n_+} E_{n_+} 
\Psi^{\dag}_{n_+}(\vec x)
\Psi_{n_+}(\vec x)+
\frac 1 2\sum\!\!\!\!\!\!\!\int _{n_+}E^{(0)}_{n_+} 
\Psi^{\dag(0)}_{n_+}(\vec x) \Psi^{(0)}
_{n_+}(\vec x).\label{sommesym}
\eea
  
The states are normalized in such a way that the two
 densities of states $\rho(E)$, with
 and without the star (free case), coincide 
 in the asymptotic region $|\vec x|\to \infty$
so that  $w(|{\vec x}| \to \infty)=0$.

For later use let us just remind the reader that the same result
 (\ref{somme})
may be obtained from  the time Fourier-transformed propagator:

\bea
\tilde S_F(\vec x,\vec y, q^0)\gamma_0 = i\ \sum\!\!\!\!\!\!\!\int\  
\frac{\Psi_{n_+}(\vec x)\Psi^{\dag}_{n_+}(\vec y)}
      {q^0-E_{n_+}+i\epsilon} +
      \frac{\Psi_{n_-}(\vec x)\Psi^{\dag}_{n_-}(\vec y)}
      {q^0-E_{n_-}-i\epsilon}.\label{fourier}
\eea

The result (\ref{somme}) is obtained from (\ref{fourier}) by integrating 
 $-q^0\,S_F(\vec x,\vec y, q^0)/(2\pi)$  on the complex $q^0$ plane, by 
 closing the contour
on the upper half. Obviously an appropriate choice for the $i\epsilon$ 
convention is crucial here, as discussed at length in \cite{note}. It is also
clear that  formula (\ref{somme}) leads to much simpler calculations than the
direct calculation of the loop in (\ref{Wanalyt}).

Obtaining the weak self-energy of the finite neutron star is equivalent to
 calculating the neutrino propagation in a background of 
neutron fields density with a border. We did that analytically in $(1\ +\ 1)$ 
dimensions with
a sharp border, and in $(3\ + \ 1)$ dimensions with a flat border.
 The spherical
symmetry has been studied both
analytically and numerically.

\section{The toy example of $(1\ + \ 1)$ dimensions}
\label{1+1}

Working directly with eq. (\ref{Wanalyt}) requires 
 an interacting Feynman propagator that takes 
the existence of a border into account.
 Doing this in $(3\  +\  1)$ dimensions is a  
big task. 

 It is feasible and theoretically fruitful  to perform the
analytical  calculation of eq. (\ref{Wanalyt}) in $(1\  +\  1)$
 dimensions as a toy model.
 The knowledge extracted from that study will help us face the 
 $(3\ +\ 1)$ realistic case by using the general formalism  anticipated in
 section \ref{general}. We will now summarize the ($1\ +\ 1$)-dimensional
results presented in more detail in \cite{note}.
 
 The ($1\ +\ 1$)-dimensional toy model is presented in ref. \cite{note}. 
  We will summarize the computation and the result in order to extract 
  useful information for the more realistic case. 

We consider  $(1\   + \ 1)$ massless fermion Feynman 
propagators in  a space with 
two regions
separated by a border. The two regions, inside and  outside the star,
 have different fermion dispersion relations,  
 as seen in eq. (\ref{disp}). 
 
 For simplicity, we consider only one sharp border located at $z=0$ and use
   an effective neutrino Lagrangian, which 
       summarizes the interaction with the neutrons \cite{Mor97}:

\beq
{\cal L}_{\mathrm{eff}}=i{\overline{\nu}} (z) \rlap/ \partial  \nu (z) - b \theta(z)
{\overline{\nu}}
 (z) \gamma^0
\nu (z) \ \  ,      \label{lagrangian}
\eeq

\noindent where $b$ is given by eq. (\ref{pmb}). The details of the computation of
the interacting propagator can be found in ref. \cite{note}. In momentum
space, the propagator can be written as follows:
 
\bea
\begin{array}{l}S_F(q^f,q^i)= 2\pi\delta\left( q_z^f-q_z^i \right) 
{i\over (\rlap/q_>)^*}  
\ + \ {b\over 2}\ {1\over q_z^f -q_z^i+i\varepsilon} \\
\times \left\{ {1\over
(\rlap/q_>^i)^*}\gamma^0{1\over \rlap/q_<^f} 
\left(1+ \mbox{sign}(q_0)\alpha_z \right) \right.  
\left. + {1\over (\rlap/q_>^f)^*}\gamma^0{1\over \rlap/q_<^i}
\left(1-\mbox{sign}(q_0)\alpha_z \right)
\right\} \ \ , \end{array}
\label{4-11}
\eea

\noindent where:

\bea 
&&{1\over \rlap/q_<}={1\over \rlap/q}= {\rlap/q_< \over q_<^2+i\varepsilon} 
\ \mbox{with} \ \ \left(q_<\right)^{\mu}=q^{\mu}=(q_0, {\vec q}) \ \ , \nn
\\ 
&&{1\over (\rlap/q_>)^*}={\rlap/q_> \over q_>^2+is\varepsilon} 
\ \mbox{with} \ \ \left(q_>\right)^{\mu}=(q_0-b, {\vec q}) \ \ ;
\label{4-12}
\eea

\noindent with $s=\mbox{sign}(q_0)\mbox{sign}(q_0-b)$,   
and $\alpha_z=\gamma^0\gamma^1$. 
The sign $s$ results from the
appropriate Feynman boundary conditions and from the right choice of
the zero energy level. It gives an adequate {\it time
convention}. We should emphasize that the
propagator (\ref{4-11}) is infrared-safe, all the $\rlap/q$'s in the
denominators being regularized by Feynman's prescription for 
the distribution
of the propagator poles in the complex-$q_0$ plane\cite{note}. 
 
 The expression for $1/(\rlap/q)^*$, given by eq. (\ref{4-12}), can be
appropriately rewritten as

\bea
{i\over (\rlap/q_>)^*} = i\left\{ {1\over \rlap/q_>}+2\pi i
\rlap/q_>\delta\left(q_>^2\right)\theta(-q_0)
\theta(q_0-b)  \right\} \  .
\label{4-14}
\eea 
 In ref. \cite{note}, we have demonstrated that the main contribution to
the weak self-energy comes from the first term of the propagator
(\ref{4-11}). We thus identified the expression given in \\
eq. (\ref{4-14}) with the one for the effective propagator, which takes
into account the ``bulk'' of the neutron star. As a matter of fact, the
effective propagator for the infinite star \cite{rescue} coincides
with  the propagator (\ref{4-14}), except for the time convention
introduced by the proper boundary conditions, responsible for the
second term in the l.h.s. of eq. (\ref{4-14}). This term is nothing else than the 
 condensate contribution, i.e. the Pauli blocking effect of the neutrinos
  trapped  into the star by the attractive potential. 
  In refs. \cite{rescue} and \cite{smirnov},
the same condensate term in the l.h.s. was introduced by
hand. Equations (\ref{4-12}) and (\ref{4-14}) give a confirmation of
the idea proposed in refs. \cite{rescue} and \cite{Mor97}: the condensate is a 
consequence of the existence of a border.

The condensate is physically understandable. As we have tuned the level of the 
Dirac
 sea (see eq. (\ref{disp})) outside the star (to the left), and as our states 
 extend over all space,
  far 
 inside the star (to the right), the level corresponds to filling a Fermi sea 
 above the bottom of the potential $b<q_0<0$. 
This obviously induces a Pauli blocking effect and eq. (\ref{disp}) anticipates this result: 
  $|b|$ is a lower bound for the momentum of the positive-$q_0$ states.

Now, with the computation of the propagator, it is easy to calculate the
weak self-energy density  $w(z)$ for this  ($1\ +\ 1$)-dimensional star by 
 following  eq. (\ref{Wanalyt}). Some details about   this rather  cumbersome 
 calculation may be found in \cite{note}. It 
is worth while emphasizing that the vanishing result (\ref{null}) for the border
contribution requires the change in the order of integration in the
momentum space over
the variables $q_0,q_z^i$ and $q_z^f$.  The latter is only
possible in the framework of an implicit  ultraviolet regularization
scheme. As a matter of
fact, the  choice of a certain order of integration over the variables
is an ultraviolet regularization method.
By integrating over the momenta we found that this difference vanishes exactly:

\bea\label{null}
w(z)=0 \ \ .
\eea

This vanishing interaction energy in $(1\ +\ 1)$ dimensions has a simple physical
explanation: the presence of the
border does not disturb the wave functions (up to a phase). There is a
one-to-one correspondence of the negative-energy states in the two media
 $z<0$ and $z>0$, and
by interchanging  the sum and integral as an ultraviolet regularization
 procedure,
 each term in 
 eq. (\ref{somme}) vanishes exactly. In
other words, massless propagating  neutrinos are not reflected by the 
border of a $(1\ +\ 1)$
 star, and the probability density is not affected by the border either.
 
The use of eq. (\ref{somme}) allows a  trivial {\it generalization of 
the vanishing result
(\ref{null}) in $(1\ +\ 1)$ dimensions to more complicated structures,
 for example 
a star with two sharp borders or, still better, to a continuously
 varying neutron 
density}, as long as the density vanishes asymptotically outside the star.

\section{Flat border in $(3\ +\ 1)$ dimensions}\label{flat}

The computation of the neutrino loops (fig. \ref{diagramatic}) in the
$(3 \ + \ 1)$-dimensional case, with a finite star, as was done in the 
preceding section for $(1\ +\ 1)$ dimensions, is technically rather difficult, 
if only because the neutrino propagators are not simple. 
In the preceding section we have seen that the use of eq. (\ref{somme})
simplified the calculation  a lot, reducing it practically to triviality in 
the latter case.  We will therefore stick to it in the $(3\ +\ 1)$ case. 

 We might have faced the $(3\ +\ 1)$-dimensional 
problem  analytically with interacting propagators and the
 approach that led to eq. (\ref{Wanalyt}), by using a simplifying 
assumption: {\it the
influence of the border can be reasonably neglected if the weak energy
density is computed far inside the star}. This amounts to considering
 a star large enough for the contribution of the
{\it bulk} effective propagator (\ref{4-14}) to be the main one. 
In other words, it would amount to assuming {\it an
infinite star} but  taking into account the existence of a matter-free
vacuum  infinitely far from the star centre. This matter-free vacuum
allows us to fix the zero energy level, and consequently the adequate
time convention for the effective propagator \cite{note}.

The latter calculation for the infinite star has indeed already been 
performed in our
 previous work \cite{rescue}. There, $w(\vec x)$ was computed using the 
 Schwinger--Dyson expansion of \\eq. (16) in \cite{rescue} and the 
 Pauli--Villars procedure
  as a means of  
 regularization. As already noted in the introduction and in ref. \cite{note},
   a pole ($q_0=|\vec q| +b$, in the case $|\vec q|<b$) has been forgotten 
   (see fig. 1 in
   \cite{note}) in 
   the analytic continuation, which led to a wrong non-null result for 
   $w(\vec x)$ 
   when the neutrino condensate effect was added by hand. 
   This happened because we did not use  the  $i \epsilon$ 
   prescription correctly: it 
     had to be imposed in the
propagators; this  was  already  corrected in ref. \cite{note}.  Taking
into account the results from refs. \cite{rescue} and \cite{note}, we can
conclude that {\it the weak self-energy is null for an infinite stationary
star.} 
It should be repeated that following  the correct $i \epsilon$ prescription,  when using 
propagators,
 is equivalent to taking the zero energy level for the states  by matching to 
 the free states 
  far outside the star. 
 Of course the definition of the zero energy level is essential when using eq.  (\ref{somme}),
  as we shall now do.

The null result for the $(3\ +\ 1)$-dimensional infinite star is a consequence 
of
neglecting the existence of border effects.
 Now, before  focusing  on the spherical problem and in order to
estimate  the border  effect, we concentrate
 on the geometrically easier problem: {\it matter-free and
neutron vacua separated by a flat border} \footnote{This problem can
be understood as the consequence of zooming on the spherical border, the
neutrinos wavelength being much smaller than the radius of the
star.}. For simplicity, the plane $z=0$ is assumed to be the flat border.  
The application of the approach resulting in eq. (\ref{somme}) requires
 a complete set of eigenfunctions for the Hamiltonian with the
flat border. 
These eigenfunctions should
be normalized following the same criteria as previously presented
in section \ref{general}:  they asymptotically behave as  plane waves  far outside
the star, in the sense that they provide, far outside the star, 
the same probability and energy density as the free plane waves 
(solutions without a star): $w(z=-\infty)=0$. 
This condition is imposed inside each 
Hilbert subspace with energy $E  \in [E,E+dE]$.

Following \cite{Gav94}, we obtain the following set of
negative energy eigenfunctions: 

\bea
\psi^{\mathrm{free}}_{n_-}(x)&=& N_{n_-} \left[ \left( e^{ik_i^<x}v_h(\vec{k}_i^<)+
e^{ik_r^<x}{\cal R}v_h(\vec{k}_i^<) \right) \ \theta(-z)\right.  \cr
&+& \left. e^{ik_t^>x}(1+{\cal R})v_h(\vec{k}_i^<) \ \theta(z) \right]  \cr
\psi^{\mathrm{mtt}}_{n_-}(x)&=&N_{n_-} \left[ e^{i\vec{k}^{<}_r x} (1+{\cal J}) v_h(\vec{k}^{>}_{mi})
\theta(-z) \right.\cr
&+& \left.  \left( e^{ik_{mi}^>x}v_h(\vec{k}_{mi}^>)+
e^{ik_t^>x}{\cal J}v_h(\vec{k}_{mi}^>) \right) \ \theta(z) \right] \ \
,
\label{Ortho}
\eea

\noindent where they are directly related to {\it
incoming} wave packets, the first coming from the {\it free} vacuum and
the second one from the {\it matter} vacuum; $n_-$ labels the negative-energy 
eigenstate,
$n_{\pm}=(\pm|E|,\vec{k}_p, h=-1)$; $\vec{k}_p$
stands for the projection of the momentum on the flat border, and $h$
is the negative helicity of standard neutrinos; $N_{n_-}$ is the appropriate
normalization factor. We also have
$k^{>}=(|E|, \vec{k}^{>})$, $k^{<}=(|E|, \vec{k}^{<})$, and

\bea
& \vec{k}_i^{<}= (k_x, k_y, k_z^<) \ , & \vec{k}_r^<= (k_x, k_y,
-k_z^<) \ ,  \\
& \vec{k}_t^>= (k_x, k_y, (k_z^>)^*) \ , & \vec{k}_{mi}^>= (k_x, k_y,
-k_z^>) \ ;
\label{Imp}
\eea

\noindent the $z$-components of the momentum are defined as
\bea
k_z^<&=&+\sqrt{E^2-\vec{k}_p^2}\cr
k_z^>&=&+\sqrt{(|E|-|b|)^2-\vec{k}_p^2}\ .
\eea
The reflection
coefficients $\cal{R}$ and $\cal{J}$ can be computed following, for
instance, the work of Gavela { et al.} \cite{Gav94}, we find:
${\cal R}=-{\cal J}=R_0 \vec{\alpha}  \hat{k}_p$, where
$\hat{k}_p$ is the unitary vector in the direction of the transverse
momentum $\vec{k}_p$,  and $\vec{\alpha}=\gamma^0 \vec{\gamma}$;
 $R_0$ is given by:

\beq
R_0= { 2 |b| |\vec{k}_p|\over (k_z^< + k_z^>)^2 - b^2} \ \ .
\label{R0}
\eeq

If we consider a neutrino outside the star, with four-momentum $(|E|,
\vec{k}^{<})$, it can easily be seen that, for $|E|-|b|<|\vec{k}_p|<|E|$, the matching
condition generates a momentum inside the star with an imaginary
component, $k_z^{>}=+i\sqrt{\vec{k}_p^2-(|E|-|b|)^2}$. Thus, a
plane wave far outside the star in the above-mentioned parameter range
generates a damped wave inside. We have, in other words, {\it a
non-penetrating wave}. This is a crucial fact, because the probability
density of waves inside and outside the star will be modified in a
different way. It will be seen that  the non-zero result for the weak
self-energy, in both the flat and the spherical border cases, is dominated by
the effect of these {\it non-penetrating waves}. 

Following Gavela { et al.}, we can be sure that the
only negative-energy {\it non-penetrating} solution, once $n$ is
fixed, is the first eigenstate of eqs. (\ref{Ortho}). Nevertheless,
when transmission occurs, both eigenstates are needed to span the
whole eigenspace. It can be seen that \\
eq. (\ref{Ortho}) appears to be
a complete set of eigenfunctions \cite{Gav94}, \cite{Rod97}.

The normalization constants are chosen such that the probability
density is asymptotically equal to the free one. For a given energy
this can only be achieved on average up to an oscillating term that 
will be considered as a negligible local fluctuation. This
normalization convention will reach its unambiguous meaning in the next
section.  There, the asymptotic density of the solutions far outside
the star is tuned to the one in the free vacuum. This asymptotic region
is taken to fix the normalization because it extends to infinity {\it
while} the star occupies a limited region. 
Applying then 
eq. (\ref{somme}),  we obtain the following weak energy density per energy:

\beq
{\partial w\over \partial |E|}(z)=-{4 \pi \over  (2 \pi)^3} |E|^3 \left\{
\Gamma(E,z)-\Gamma^0(E,z)\right\} \ , 
\label{density}
\eeq

\noindent where $\Gamma^0(E, z)=1$, for free plane waves; 
$\Gamma(E, z)$, for instance in the region $|E|>|b|$, can be written as:

\beq
\Gamma(E, z)=\theta(-z) \ + \ \gamma(E) \theta(z) \ \ ,
\label{Gamma}
\eeq

\noindent where

\beq
\gamma(E)=\int_0^{1-|b^*|}dx {x\over \sqrt{1-x^2}}\ {1+ {R^*_0\over
1+(R^*_0)^2}x \over 1- {R^*_0\over 1+(R^*_0)^2} {x\over 1-|b^*|}} \ \ .
\label{gamma}
\eeq

\noindent The parameter $b^*$ is the dimensionless quantity $b/|E|$, and
$R^*_0$ is the same function $R_0$ as given by eq. (\ref{R0}), expressed in
terms of the variable $x=|\vec{k}_p|/ |E|$ and of the parameter $b^*$. It
is easy to see that $R_0$ is a real quantity, except for the
non-penetrating waves, the result $|R_0|^2=1$ being satisfied in the
latter case. It is important to insist on the fact that the local fluctuations
originated by the interference of incident and reflected waves have
been neglected in both regions, $z>0$ and $z<0$. The fact that these fluctuations
contribute in a negligible way can be easily seen by performing the
appropriate normalization of the eigenstates in a certain box. The
contribution of the damped waves inside the star has also been 
neglected. That is why we integrate only for penetrating waves to
obtain the function $\gamma(E)$, i.e. the reason of the upper bound in
eq. (\ref{gamma}), which implies that $R_0$ is always real
in the l.h.s. of eq. (\ref{gamma}). 
The latter damped waves occupy a thin layer inside the star and are 
necessary to make  the probability density on the border continuous.
 A similar 
effect will be discussed  in the following section for the spherical case.

In order to show the dominance of the effect of the non-penetrating
waves, which is the main aim of this planar computation, the weak energy
density (\ref{density}) will be compared with a naive
estimate of the one coming from non-penetrating waves: 
integrating  simply the plane wave density  in the two regions summed 
 over all the allowed momenta in both 
$k_z=\sqrt{E^2-|\vec{k}_p|^2}$ for $z<0$ and
$k_z=\sqrt{(|E|-|b|)^2-|\vec{k}_p|^2}$ for $z<0$, we obtain

\beq
\gamma^S(E)=\left( {|E|-|b|\over |E|}\right)^2 \ .
\label{gam2}
\eeq

\noindent Equation (\ref{gam2}) results
from computing separately the density of states in the star
and in the free vacua, the eigenstates in these vacua being obtained for
translationally invariant Hamiltonians, which {\it do not generate
reflection on the border}. The difference between $\gamma^S(E)$ and 
$\gamma(E)$ is precisely due to this neglecting of reflection. 
The result (\ref{gam2}) only  accounts
for the fact that certain eigenstates, being allowed in the free
vacuum, are not inside the star. There is an obvious {\it one-to-one}
correspondence between these eigenstates and the ones for the
untranslational-invariant Hamiltonian (\ref{lagrangian}), which we called
non-penetrating waves.

\begin{figure}
\begin{center}
\mbox{\epsfig{file=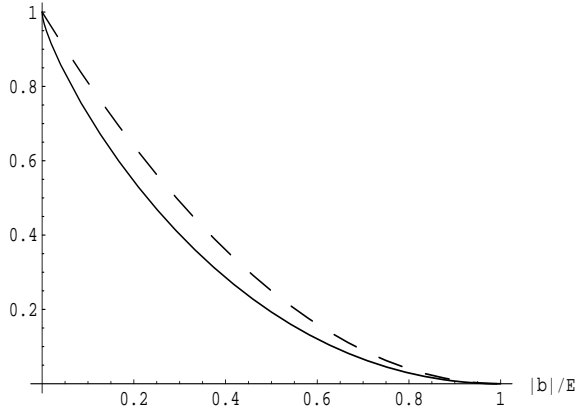,height=8cm}}
\end{center}\vskip -1.5cm
\caption{{\it {Plots of $\gamma(E)$ (dashed line) and $\gamma^S(E)$ 
(solid line), defined
in the text, as a function of $|b|/|E|$.}}}
\label{Comp}
\end{figure}

\noindent In fig. \ref{Comp}, the functions $\gamma(E)$ and
$\gamma^S(E)$ have been plotted. 

The small difference between the two curves in fig. \ref{Comp}
shows that the dominant effect of the border comes from the non-penetrating
waves. This result will be confirmed in the next section. A few comments are
 in
order here. First, the new feature in the $(3\ +\ 1)$-dimensional case with 
respect 
to the $(1\ +\ 1)$ one is that the non-trivial neutrino
 {\it refraction index not only bends the
penetrating waves and induces a reflection, but also, beyond the limiting angle,
induces total reflection}. This well-known phenomenon, acting here on the negative
energy states, induces the dominant contribution to the star weak self-energy
density. It must be stressed that this effect of non-penetrating waves
is utterly unrelated to any Pauli blocking effect from the neutrino condensate
inside the star. Indeed, as we have shown in the $(1\ +\ 1)$ case, in which the
condensate exists, its effect is precisely to equate the energy density inside
the star to the one outside. On the other hand, the non-penetrating waves are
reflected, not because of some states, which are already occupied, but because
 they
tend to occupy states that simply do not exist, with an imaginary momentum.
 Finally, let us insist that \\eqs. (\ref{density}) and (\ref{gamma}) imply 
 that 
{\it the border effect discovered here is a volume effect}, i.e. it affects
 the energy density by an almost constant amount in the whole volume occupied 
 by the star. This result is rather unexpected, a border effect being 
 thought to act only on the surface.  Of course it relies on the hypothesis
  that the wave packets extend coherently over the whole star. As will be 
  discussed in section \ref{discussion}, such a hypothesis is perfectly
   sound for low-energy neutrinos.

\section{Realistic star: spherical 3-d}\label{spherical}

We study  a massless neutrino in the presence of an external symmetric
 static
electroweak
potential of finite range due to the interaction with the neutrons of 
the star. 
In order to calculate the weak self-energy 
$\int d^3 r \ w(\vec r) =\int d^3 r 
\ (\langle\hat{0}|{\cal H}(0)|{\hat{0}}\rangle - 
\langle{0}|{\cal H}_0(0)|{0}\rangle)$, 
which is nothing else than the difference 
(\ref{somme}) integrated over space, we need 
to use the spherical Bessel functions \cite{watson} as a basis for the 
solutions of
the Dirac equation.  The effect of the static neutrons is summarized 
in the
 spherically symmetric  square-well potential of depth $b<0$ 
 ($b\sim -20$ eV): 
\bea
V(r)=\left\{\begin{array}{ll}\ b <0 & {\mathrm{for}}\ \  r\leq R\\
\  0 & {\mathrm{for }}\ \  r > R\, \end{array}\right.
\label{potential}
\eea
$R$ being the radius of the star. 
\\
>From the effective
Lagrangian of eq. (\ref{lageff}), 
\bea
{\cal {L}}_{\mathrm{eff}}={\overline{\psi}}
\left(i\partial\!\!\!/ -
V(r)\gamma_0\right)\ \psi ( r) 
 \ , \label{lag-spher}
\eea
the Dirac equation is:
\bea
{\vec\alpha} \cdot  {\vec p}\ \psi({\vec r})=
\left (E -V(r)\right )\psi({\vec r})\ ,
\label{Dirac}
\eea
\noindent where $$\vec{\bf\alpha} =
\left (\begin{array}{cc}0&\vec{\bf\sigma} \\ 
\vec{\bf\sigma}&0\end{array}\right ) \ .$$
Turning the kinetic energy operator $\vec{\alpha} \cdot {\vec p}$ into 
spherical polar coordinates, we obtain the eigensolutions :
\bea
\psi_{\kappa,m}({\vec r})=N\ \left ( \begin{array}{l}
\ u_1(r)\ 
\chi_\kappa^m(\theta,\phi) \\i u_2(r)\ 
\chi_{-\kappa}^m(\theta,\phi)
\end{array}\right )\ ,
\label{psi}
\eea
\noindent where $N$ is a normalization factor and 
\noindent $\chi_{\pm \kappa}^m(\theta,\phi)$ are the two-spinors \cite{greiner}
written in terms of the spherical harmonics $ Y_l^m$ and spin $1/2$ 
eigensolutions $\chi_{{1\over 2},\mu_s}$:

   \bea
  \chi_{\kappa>0}^m(\theta,\phi)=
  {\displaystyle{\sum_{\mu,\mu_s}\left\langle \kappa \mu \frac 1 2
   \mu_s{\bigg{|}}\left(\kappa -\frac 1 2 \right) m\right\rangle
  { Y}_\kappa^\mu(\theta,\phi)\chi_{{1\over 2},\mu_s}}}=\,
   \left ( \begin{array}{c}
-\sqrt{\frac {\kappa-m+\frac 1 2}{2\kappa+1}}Y_{\kappa,m-\frac 1 2}\\ 
\sqrt{\frac {\kappa+m+\frac 1 2}{2\kappa+1}}Y_{\kappa,m+\frac 1 2}
\end{array}\right ),\label{definition+}
  \eea 
  \bea
 \!\! \chi_{\kappa<0}^m(\theta,\phi)\!=\!\sum_{\mu,\mu_s}\left\langle (-\kappa-1) 
 \mu \frac 1 2
   \mu_s{\bigg{|}}\left(-\kappa-\frac 1 2\right ) m\right\rangle\!
  { Y}_{-\kappa-1}^\mu(\theta,\phi)\chi_{{1\over 2},\mu_s}\!=\!
   \left (\!\begin{array}{c}
\sqrt{\frac {-\kappa+m-\frac 1 2}{-2\kappa-1}}Y_{-\kappa-1,m-\frac 1 2}\\ 
\sqrt{\frac {-\kappa-m-\frac 1 2}{-2\kappa-1}}Y_{-\kappa-1,m+\frac 1 2}
\end{array}\!\right ) \label{definition-}
  \eea

Here $\chi_{ \pm |\kappa|}^m(\theta,\phi)$ are eigensolutions of $J^2$, $J_z$,  $L^2$
 and $S^2$, where $\vec J=\vec L+\vec S$, $J_z$, $\vec L$ and $\vec S$ are
 respectively the total angular momentum, its projection along the $z$-axis,
 the orbital angular momentum and the spin angular momentum 
verifying the eigenvalue equations:

\bea
 \left \{ \begin{array}{l}( \vec{\bf\sigma} \cdot \vec { L} + 1 )
\ \chi_{\pm |\kappa|}^m(\theta,\phi)\ =\ \mp |\kappa|
 \chi_{\pm |\kappa|}^m(\theta,\phi)
\ , \\
  {J_z}\chi_{\pm |\kappa|}^m(\theta,\phi)\ =\ 
  m \chi_{\pm |\kappa|}^m(\theta,\phi)\ ,\ \  -J\leq m\leq +J\ ,
 \\
 |\kappa|=(1, 2, ...)\ , \ \ 
 |\kappa|=J + 1/2\ .\end{array}\right.
\label{kappa}
\eea

\noindent With this notation, we can easily check that the two chirality
states are described by the same spinor of eq. (\ref{psi}):
\bea
\psi^m_{\kappa_{L,R}}={\frac 1 2}\ (1\ \mp\ \gamma_5\ )\psi^m_{|\kappa|}\ ,
\label{chiralite}
\eea
\noindent the positive and the negative values of $\kappa$ being related 
through the relation:
\bea {\displaystyle{i\ \gamma_5\ \psi^m_{-|\kappa|}=\psi^m_{|\kappa|}}}\ .
\label{psi-L-R}
\eea

It then suffices to compute $\psi^m_{\kappa}$ for positive
 values of $\kappa$, use eq. (\ref{psi-L-R}) for the negative values of
 $\kappa$ and then get the left-chirality neutrino state using eq. 
 (\ref{chiralite}). Let us from now on write the positive $\kappa$'s
  as $\kappa\equiv l=1,\ 2,\ 3,....$, with $J=l-1/2$ 
  (see eq. (\ref{definition+})).

 The radial functions $u_{1,2}$ defined in eq. (\ref{psi})
 verify the following coupled differential equations:
 \bea
{d\over d r}\left ( \begin{array}{l}
r\  u_1(r) \\ r\ u_2(r)\end{array}\right )=\left [ \begin{array}{cc}
\ {-l\over r} & (E-V(r)) \\
- (E-V(r)) & {l\over r} \end{array}\right ]\ \ \left ( \begin{array}{l}
r \ u_1(r) \\ r\  u_2(r)\end{array}\right )\ .
\label{diffequa}
\eea

Now, decoupling the equations and using eq. (\ref{potential}) and the fact that 
the
solution must be regular in the centre of the core, i.e. $r\to 0$, we get:
\bea
\left \{ \begin{array}{l}
u_1(r)\ =\alpha\big  [  \  j_l\left (k\ r\right )\ 
 \theta(R-r) + (B\  j_l (E\ r) + C\  y_l(E\
 r))\ \theta(r-R) \big  ]\\
 u_2(r)\ =\alpha\big  [\   j_{l -1}\left(k\ r\right)\ 
  \theta(R-r) + 
 (B\  j_{l -1} (E\ r) + C\  y_{l -1}(E\
 r))\ \theta(r-R)\big  ] \ ,\end{array}  
\right.
 \label{solution}
\eea
\noindent where we define 
\bea
k=E-b ,\nn \\
\alpha={1\over \sqrt{B^2+C^2}}\ ;\label{alpha}
\eea
 $B,\ C$ being integration
 coefficients to be fixed from 
the matching condition at the surface. The factor $\alpha$ 
will be justified  below to guarantee the good asymptotic behaviour of
functions $u_{1,2}(r)$. 
 The functions 
$j_{l}$ and $y_{l}$ are the spherical Bessel functions of the first 
and
second kind. 
 In the inner region, the solution can be written only in terms of $j_l$
 because the wave functions have to be regular at the origin and
  the
$y_l(\rho)$'s are not 
(see eq. (\ref{comportement})).

According to eq. (\ref{somme}), we are going to use only negative energy states
and thus negative arguments for these Bessel functions ($E,k<0$) 
 and the following relations are helpful:
 \bea
 j_l(\rho)= (-1)^{l+1} j_l(-\rho)\  ,\qquad 
 y_l(\rho)= (-1)^{l} y_l(-\rho)\ .
 \label{parity}
 \eea

 The solutions $u_{1,2}$ have been written in the form (\ref{solution}) in
 order to match, up to a phase, the free solutions as $r\to \infty$, 
 where the asymptotic
 behaviour of the Bessel
 functions is: 
 \bea
 j_l(\rho){\displaystyle{\ \stackrel{\sim}{{}_{\rho\to \infty}}}}
 \ \sin \left(\rho- \frac { l \pi}{2}\right){\bigg{/}}\rho\ ,\qquad 
  y_l(\rho){\displaystyle{\ \stackrel{\sim}{{}_{\rho\to \infty}}}}\ 
  \cos \left(\rho- \frac { l \pi}{2}\right){\bigg{/}}\rho\ .
 \label{comportinfinity}
 \eea
This behaviour implies the one followed by our solutions $u_{1,2}$:
\bea
u_1(E\  r)&{\displaystyle{\ \stackrel{\sim}{{}_{r\to \infty}}}}
 &\ \sin \left(E \  r- \frac { \pi l}{2}+\phi\right){\bigg{/}} E\  r\ ,\cr
u_2(E\  r)&{\displaystyle{\ \stackrel{\sim}{{}_{ r\to \infty}}}}&\ 
  \cos \left (E\  r- \frac { \pi l}{2}+\phi\right){\bigg{/}}E\  r\ ,\cr
{\mathrm{where}}\ \qquad\qquad\qquad \ \phi  &=& \arctan (C/B)\ .
\eea
  Needless to say that in the absence of the star ($b=0$),
  $\alpha =1$
  and the spherical solutions fit exactly the free ones, in the asymptotic
  region  $r\to\infty$.
  
Both inner and outer solutions must join at $r=R$, and this 
matching fixes the 
coefficients
$B$ and $C$ as follows:
\bea
 \begin{array}{l}
B\ =\  {\displaystyle{{ j_l (k\ R)\ y_{l -1}(E\ R) -  
j_{l -1}(k\ R)\  y_l (E\ R) \over
 j_l (E\ R)\ y_{l -1}(E\ R) -  
j_{l -1}(E\ R)\  y_l (E\ R)}}}\ ,\\ \\
 C\  =\  {\displaystyle{ { j_{l -1} (k\ R)\ j_l(E\ R)-  
 j_{l -1} (E\ R)\ j_l (k\ R) \over
 j_l(E\ R)\ y_{l -1}(E\ R) -  
j_{l -1}(E\ R)\  y_l (E\ R)}}}\ .
  \end{array} 
 \label{bessel}
\eea

Finally, in order to fix the solutions definitively, we need to calculate the 
normalization factor $N$ by following the asymptotic
 normalization convention already introduced in the previous section: 
we impose the
probability density to be asymptotically  equal to the free one. Since 
we have 
fitted $\alpha$ to 
make the solutions with $b\neq 0$
coincide, up to a phase, with those of the free case ($b=0$)
 when $r\to \infty$, we need to compute $N$ only for  the free case. 
In the latter case, $b=0, \alpha=1$, the Hilbert space of spherical
 solutions is the one of
the free plane  waves ${1\over (2\pi)^3}e^{i\vec p\cdot\vec r}$.
 The  solutions  
(see eq. (\ref{bessel})) reduce to 

\beq
\psi^m_{l}({\bf r})=N\ \left ( \begin{array}{l}
 j_l (E\ r)\ 
\chi_l^m(\theta,\phi) \\i j_{l-1} (E\ r) \ 
\chi_{-l}^m(\theta,\phi)
\end{array}\right )\  .
\eeq
Now 
 we compute the density of these states $\rho(E)$ in the Hilbert subspace
corresponding to\\
 $E\in [E,E+dE]$, by summing up all the quantum numbers and the
angular degrees of freedom, while keeping the left chirality and spin fixed
because we are 
 counting the negative-energy and left-handed neutrino states. 
 This gives  the probability $\rho(E)dE={4\pi
 E^2\over (2\pi)^3} dE$. Using the relation \cite{watson} 
\bea\sum_{\mu=0}^\infty (2\mu+1)|j_\mu|^2
 =1\ ,\label{identity}\eea 
\noindent we obtain: 
\bea
N=E/\sqrt{2\pi}\ .
\label{norm}
\eea

Now that the solutions of the Dirac equation are known, we can 
calculate numerically $w(r)$, which can be
 written  following eq. (\ref{somme}) as  \cite{schwinger}:
\bea
w(\vec r)= \int dE\ \sum_{l,m} E\ \left(\psi^m_{l}\dag (\vec r)\ \psi^m_{l}(\vec r)-
 \psi^{m\ 
  (0)}_{l}\dag (\vec r)\ \psi^{m\ (0)}_{l}(\vec r)\right )\ ,
\eea
\noindent where the sum is over all the degrees of freedom of the 
negative-energy left-handed neutrinos and where the  subscript $(0)$ refers to 
the matter-free 
solutions ($b=0$).
 The integration over the angular part and the summation over $m$ is direct, 
  using eqs. (\ref{definition+}) and (\ref{definition-}), which lead to \bea  
{\frac{ 1} {4 \pi}}\int \ d\Omega \sum_{m}
\chi\dag_\l^m(\theta,\phi)\chi_\l^m(\theta,\phi)={(2J+1)\over 4\pi}={2l\over 4 \pi}\ .
\eea 
\noindent The weak self-energy density ($w(r)\equiv dW/dV=
{{{dW\over 4\pi r^2 dr}}}$,
$W$ being the
total weak energy and $V$ the volume) is then simplified to
 (using the spherical symmetry);
 \bea
 w(r)=\int^0_{-\infty} dE\  {E^3 \over 2 \pi^2}
  \sum_{l}l(\rho_l(k,r)-\rho^{(0)}_l(E,r))\ ,\label{sphere}
 \eea
\noindent where we have defined
\bea\begin{array}{l}
\rho_l(k,r)=\alpha^2 \left \{\left [j_l^2(k \ r)+j_{l-1}^2(k \ r)\right ]
\theta(R-r)\right.+\\
\left.\quad\quad\quad\ \ \ \left [(B j_l(E\  r)+C y_{l}(E \ r))^2
+ (B j_{l-1}(E\ r)+C y_{l-1}(E\  r))^2\right ]\theta(r-R)\right\} \ , \\ 
\rho_l^{(0)}(E,r)=j_l^2(E\  r)+j_{l-1}^2(E \ r)\ ,\end{array}
\label{densites}
\eea
\noindent  $k$ and  $\alpha$ being given by eq. (\ref{alpha}).
We were not able to perform the summation over all the values $l$ 
and negative energies 
  analytically. We
then use  numerical computations of the Bessel functions, 
and use some knowledge (see ref. \cite{watson}) about these 
 functions in order to justify the 
truncation of  
these infinite summations in $l$, as will be explained now.

Let us define the function 
\bea
f(E,r,l_2)=\sum_{l=1}^{l_2}l(\rho_l(k,r)-\rho^{(0)}_l(E,r))\ ,\label{function}
\eea

We  have observed that for $r > R$  and fixed values of $E$ and 
$ l_{2}$, the result becomes rather stable from  $r\simgt R$ up to $r\gg R$,
while  there is a sudden change  when we pass 
 the border $r=R$. When we integrate the density over $r$, the main
 contribution to (\ref{function}) comes from inside  the star, $r\in [0,R]$ 
 and thus
  the effect is grossly proportional to the volume.
  
  In order to understand these results  qualitatively, a few remarks about the
   behaviour of Bessel functions are appropriate. The spherical Bessel functions
    are solutions of the differential equation:
	\bea
 \left({d^2\over d \rho^2}+k^2 -{l(l+ 1)\over \rho^2}\right)(\rho j_l)=0,\qquad 
 \left({d^2\over d \rho^2}+k^2 -{l(l+ 1)\over \rho^2}\right)(\rho y_l)=0\ .
 \label{equadiff}
 \eea

 For $\rho$ large, say $\rho\sim \rho_0$, we may get a hint by replacing 
 $1/\rho^2$ by $1/\rho_0^2$ 
 in eq. (\ref{equadiff}).\\
  For ${\displaystyle{k^2 -{l(l+ 1)/\rho_0^2}<0}}$, i.e.
  $l\simgt |k| \ \rho_0$, the solution  $j_l$ ($y_l$) is a damped (exploding)
   exponential
   \footnote{This damping of $j_l$ expresses the semi-classical 
fact that $l\sim ||\vec \rho_0\times \vec k||< \rho_0| k|$, the waves
corresponding to $l> \rho_0\ |k|$ are
damped.}.  This damping (explosion) is expressed by the well-known  behaviour
 of the spherical Bessel functions $j_l$ and $y_l$ in the
 vicinity of the centre:
 
 \bea
 j_l(\rho){\displaystyle{\stackrel{\sim}{{}_{\rho\to 0}}}}{\rho^l\over(2l+1)!!}\ 
 \ ,\qquad
  y_l(\rho){\displaystyle{\stackrel{\sim}{{}_{\rho\to 0}}}}
  {(2l+1)!!\over (2l+1)}\rho^{-(l+1)}\ .
    \label{comportement}
 \eea

For ${\displaystyle{k^2 -{l(l+ 1)/\rho_0^2}>0}}$, - i.e.
  $l\simlt |k| \ \rho_0$, the solution is an oscillating function as
   expressed by the asymptotic behaviour (\ref{comportinfinity}).

In practice the transition between these two asymptotic regimes is rather fast. 
In other words, the Bessel function  $j_l(\rho)$ 
 increases with $\rho$ at first as a power of $\rho$,  then increases 
 exponentially until a transition at 
$\rho=\sqrt{l(l+1)}$ where it becomes 
 an oscillating function, see eq. (\ref{comportinfinity}).
  Spherical Bessel functions are not so different from 
 \newpage
\bea
 j_l^{ap}(\rho)&=&\ \theta(\rho-\sqrt{l(l+1)})\,
 \sin \left(\rho- \frac {l \pi}{2}\right){\bigg{/}}\rho,\quad
  \forall  \rho,\cr
  y_l^{ap}(\rho)&=&\ 
  \theta(\rho-\sqrt{l(l+1)})\,\cos \left(\rho- \frac {l \pi}{2}\right){\bigg{/}}\rho,\quad 
    \rho-\sqrt{l(l+1)}>0.
 \label{appr}
 \eea 
\noindent  The exploding part of $y_l$ is not described by 
  eq. (\ref{appr}), but we do not care since, precisely from imposing
   regularity at the origin, the $y_l$'s only contribute to our 
   solutions outside the star, eq. (\ref{solution}), 
  where they are in their oscillating regime, as can be  easily checked.  
   
  In the following we  will use this approximation to guess the results and we
   will check numerically these guesses and estimate the corrections to these
    approximations. Such a strategy is  necessary since  we are totally 
    unable to calculate numerically with the realistic value of $|b|R$, which is 
    of the order  of $\sim 10^{12}$. 
This would need computing up to angular momenta larger than  $10^{12}$ !  
In practice we have modestly limited ourselves to 
$|b|R \simeq 10$ and $|E|/b$
going up to $\simeq 100$. It is then {\it mandatory to get a qualitative 
understanding of the results to be able to extrapolate them to the realistic 
values}.
Before going further, it is worth remarking that the function $j_l^{ap}$ 
defined in (\ref{appr}) verifies the saturation relation 
(\ref{identity}) at leading order in $1/(|k|r)$.

~From (\ref{appr}) it seems that the summations on $l$ may be truncated
at $l \simgt |E|r$.  In order to check this conjecture we will use 
 the free case in
 which we know the exact solution. For the plane waves, the energy density $w_p$ due to left-handed
 negative-energy neutrinos is 
\bea
w_p(r)=\int dE\  \frac{E^3} {2 \pi^2} = \int dE\, \frac{E^3} {2
\pi^2}\,\sum_{l=1}^\infty l\rho^{(0)}_l(E,r)\ , 
 \label{plan}
\eea
as can be directly computed from the plane wave functions, and also via the 
definition of $\rho^{(0)}_l$ in (\ref{densites}) with the help of 
 relation (\ref{identity}). 
The integration over $E$ in eqs. (\ref{sphere}) and (\ref{plan}) is  of course
divergent. We will come  to this ultraviolet 
problem later, in section \ref{discussion}. 

We have computed 
\bea {dG_p\over dE}\equiv {\displaystyle{\int_0^R}}
 \ 4\pi r^2 \ dr  \frac{E^3} {2
\pi^2}\label{ii}\eea
 and compared it with \bea {dG\over dE}\equiv{\displaystyle{\int_0^R}} \ 
 4\pi r^2 \ dr  \frac{E^3} {2
\pi^2} \sum_{l=1}^{l_2}l \rho^{(0)}_l(E,r)\label{iil}\eea 
in order to see the convergence of
the summation. It is confirmed that the summation 
${{\sum_{l=1}^{l_2}}}l \rho^{(0)}_l(E,r)$ is already very close to 
 its limit $1$ as soon as $l_2> |E|\ R$.
 This can be seen in fig. 3,  where we have plotted the absolute values
 of eqs. (\ref{ii}) and (\ref{iil}).

\begin{figure}[h]
\begin{center}
\mbox{\epsfig{file=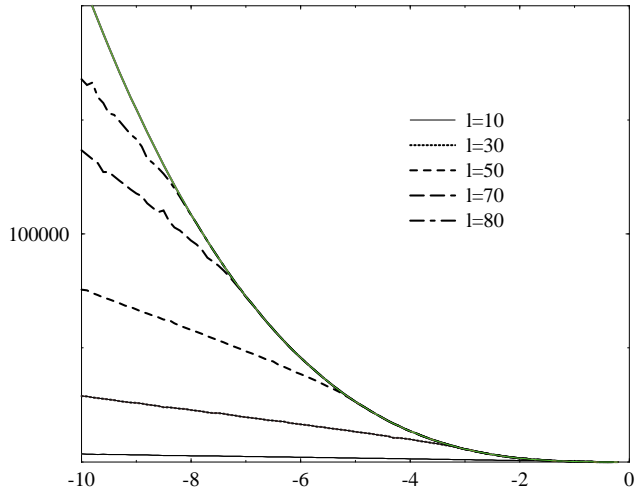,height=8cm}}
\caption{{\small \it{${\left |dG\over dE\right |}$ in eq. (\ref{iil}) 
as a function of the 
energy expressed in units of $|b|$. The summation on  $l$ goes
from 1 to the value indicated on the figure, $r$ is 
integrated between $0$ and $R$  (the radius $R$ of the star has
been fixed to $10|b^{-1}|$). The bold line represents the plane wave
density  
${\left |dG_{p}\over dE\right |}$  in eq. (\ref{ii}).}}}
\end{center}
\label{comp-4}
\end{figure}

  We have also checked that the summation in (\ref{function}) saturates when 
 $|E|\ r <l_2$.

The numerical computations show that $\alpha$, eq. (\ref{alpha}), is 
 quasi-independent of $l$ for \\$l < |E-b|\ R$ and is verified for:
\bea
\alpha^2  \stackrel{\large{\simeq}}{{}_{|E-b| R > l}} \alpha_t^2\equiv \left 
( {E-b\over E }\right )^2\, .
\label{1/astheo}
\eea

This result may be understood  by the following argument. 
Using the approximation (\ref{appr}), 
 the waves  that do not vanish inside the star near the border
  have the form of a sine divided by $(k \ R)$. 
  The corresponding ones, outside the star, are  a combination of a sine and a 
   cosine 
  divided by $(E\ R)$. Taking  the trigonometric functions  to be 
  of the same size on average, an obvious factor of 
$\alpha \simeq \alpha_t= |k/E|$ is needed for the matching at $R$ of the inside 
waves and the outside ones. 
Interestingly enough, this result leads, 
combined with (\ref{identity}), i.e. summing over all angular momenta,  
to an average energy density inside the star of 
\bea
E (E-b)^2/(2\pi^2)\ '
\label{win}\eea
which is the energy multiplying the  plane wave probability density, see eq.
 (\ref{plan}),
 but for a plane wave shifted by the potential $b$. This is
reminiscent of
 eq. (\ref{gam2}).

\begin{figure}[h]   
\begin{center}
\hskip 2.5cm
\mbox{\epsfig{file=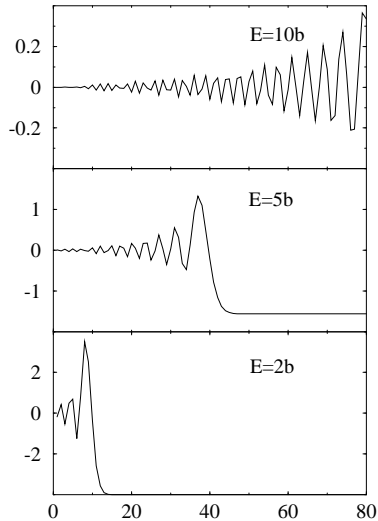,height=8cm}}
\caption[]{\it{$\alpha^2-\alpha^2_t$ (see eqs. (\ref{1/astheo}) and
(\ref{alpha})) as a function of   $l$, for different values of the energy. 
The star radius is $R=10 b^{-1}$.}} 
 \protect\label{comp0}
\end{center}
\end{figure}

 The validity of eq. (\ref{1/astheo}) is shown in fig. \ref{comp0}, where
 we compare $\alpha$  to $\alpha_t$ for different values of the energy and the
 parameter $l$. We can see on this figure that $\alpha^2-\alpha^2_t$ is 
 negligible in average,  as long as the momentum $l$
 verifies $l\ll |E-b|\ R$.   
 
 At this point, we now have  a ``naive'' or ``crude"  formula for $w(r)$: 
\bea
w_{crude}(r)\equiv\int dE\  {E\over 2\pi^2} \left( (E-b)^2-E^2 \right )\,\theta(R-r)\ , 
\label{naive}
\eea
\noindent of course, this is a first approximation, the same
as  we had reached in the flat case (see section \ref{flat}) from the function
 $\gamma^S(E)$,
eq. (\ref{gam2}). 

The reason for this agreement between the flat case and the spherical one is
 easy to 
understand physically. It is related to the non-penetrating waves. 
Let us first consider the waves with $l< |E-b|\ R$. 
We have already argued that 
they are  trigonometric functions divided by $|E-b|\ R$ 
($|E| \ R$) 
inside (outside) the star. They are matched on the border. From that matching 
itself and their oscillating behaviour, their density has to match up to 
a large 
distance from the border.
 A different and crucial effect comes from the angular momenta 
\bea
|(E - b)|\ R< l< |E| \ R \ .\label{interm}
\eea

In the  trigonometric regime these waves are functions divided by $|E|\ R$
 outside,
 while inside they are in the fast decreasing regime. The matching adjusts them for $r=R$,
  but, as $r$ decreases  inside the star, they fall very fast. 
  Essentially they are non-penetrating waves exactly analogous to the ones with 
   $|E-b|<k_p<|E|$ in the previous section. Indeed, since
    $\vec L \simeq \vec r \times \vec k$,
and noticing that the radial vector being perpendicular to the surface,
 its cross product with $\vec k$ is parallel to the surface, it becomes obvious that the 
 inequality $|E-b|<k_p<|E|$ is equivalent to eq. (\ref{interm}). \\
 
{\it We have thus seen qualitatively, and checked numerically, that the dominant 
contribution to $w(\vec r)$  is due to non-penetrating waves in the spherical case as 
well as in the planar case}.

\vskip 0.5cm

 The numerical check consisted in the following steps.\\
  In the  beginning, we compared: 
 \bea
 {dW\over dE}\equiv{\displaystyle{\int_0^{r_{max}}}}\ 4\pi r^2 dr\  {E^3 \over 2 \pi^2}
  \sum_{l=1}^{l_2}l(\rho_l(k,r)-\rho^{(0)}_l(E,r))\ , \label{df}
  \eea
for several values of $l_2$, $r_{max}$ being greater than the star radius $R$, 
with  
\bea
{dW_{crude}\over dE}\equiv{E\over 2\pi^2} 
\left( (E-b)^2-E^2 \right )\ 4\pi\ R^3/3\ .\label{dfcrude}
\eea

In fig. 5, we have plotted eq. (\ref{df})  
  as a function of the energy for several values of $l_2$ and compared
  with eq. (\ref{dfcrude}). 
  The integration over $r$ in eq. (\ref{df}) has been
 done  from $0$ to $r_{max}=3 R$. This figure confirms that, as long as 
  $ |E|\ R < l_2$,
 there is no sizeable difference between the exact result
  (\ref{df}) and the ``crude'' one (\ref{dfcrude}).
 For higher values of $ |E|\ R$, we must increase $l_2$
 in order to take all the contributions into account.
\begin{figure}[h]
\begin{center}
\vskip -2cm
\hskip 1.5cm

\mbox{\epsfig{file=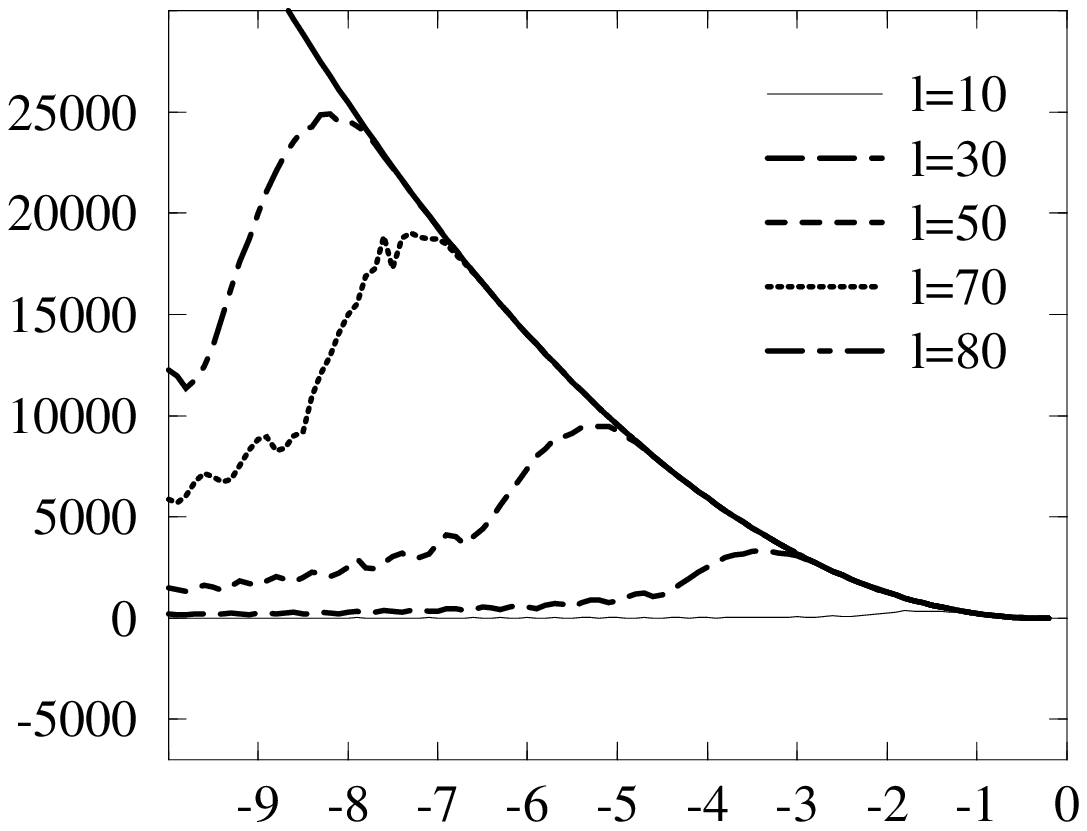,height=8cm}}
\caption{{\small \it{${\left |dW\over dE\right |}$ in eq. (\ref{df}) 
as a function of the 
energy  expressed in units of $|b|$. The summation on  $l$ goes
from 1 to the value indicated in the figure, $r$ is 
integrated between $0$ and $r_{\mathrm{max}}=30 \ |b^{-1}|$  (the radius of the star has
been fixed to $10|b^{-1}|$). The bold line represents the energy 
density 
  ${\left |dW_{crude}\over dE\right |}$  in eq. (\ref{dfcrude}).}}}
\end{center}
\label{comp1}
\end{figure}

After that, in order to check the corrections to our  
formula eq. (\ref{naive}) more accurately,  we  
 computed:
 
 \bea   
\begin{array}{l}
\qquad{\displaystyle{ {d\delta(r)\over dE}}}={\displaystyle{{d\delta_{in}\over dE}}}+
 {\displaystyle{{d\delta_{out}\over
dE}}}\label{delta}\\
 {\mathrm{where}}  \\ 
 \qquad{\displaystyle{{d\delta_{in}(r)\over dE}}}
\equiv  4\pi r^2 
  {E^3 \over 2 \pi^2}
 \left[{\displaystyle{\sum_{l=1}^{l_2}}}
 \left(\alpha^2-\alpha_t^2\right) l\left (j_l^2(k \ r)+j_{l-1}^2(k \ r)
\right )\right ]\theta(R-r)\label{in(r)}\\
{\mathrm{and}}  \\
 \qquad {\displaystyle{{d\delta_{out}(r)\over dE}}}\equiv 4\pi r^2
 {E^3 \over 2 \pi^2}\ {\displaystyle{\sum_{l=1}^{l_2}}} \ l 
 \left [
  \alpha^2  \left (
  (B j_l(E r)+C y_{l}(Er))^2
  \right.
  \right.\\
+ \left.\left.(B j_{l-1}(Er)+C y_{l-1}(Er))^2\right ) 
 - \rho^{(0)}(E,r)\right ]\theta(r-R)\ .\label{out(r)}
 \end{array}
 \eea

\noindent  Here, we are subtracting from the exact 
 formula  the
``crude'' one, expanded in Bessel functions with the help of the constant 
factor  $\alpha_t$. Indeed, from the plane wave expansion (\ref{plan}) 
and the definition of $\alpha_t^2$ (\ref{1/astheo}), it is obvious that
 the term proportional to  $\alpha_t^2$ 
in (\ref{out(r)}) sums up to the one proportional to $E(E-b)^2$ in 
(\ref{dfcrude}) and that the term proportional to $E^3$ cancels the sum of 
the terms $\rho_l^{(0)}$. We have expanded the ``crude'' contribution into 
Bessel functions in order to
 minimize the error due to the truncation of the sum.

\begin{figure}[h]   
\begin{center}
\mbox{\epsfig{file=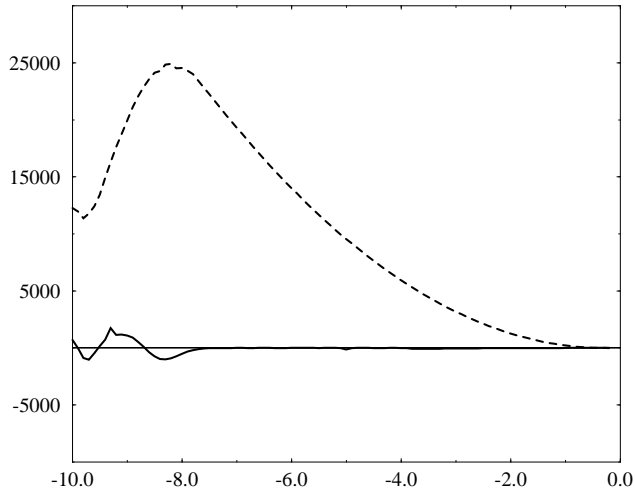,height=8cm}}
\caption[]{\it{The dashed line represents 
 ${\left |dW\over dE\right |}$ in eq. (\ref{df}) and  the solid line  
 ${\left |d\Delta\over dE\right |}$ in eq. (\ref{den}).
They are functions of the energy  expressed in units of $|b|$. The
summation over $l$ was done from $1$ to $80$ and the integration over $r$
from $r=0$ to $r=30 |b^{-1}|$ (the star radius being $R= 10 |b^{-1}|$).
 }} \protect\label{comp2-1}
\end{center}
\end{figure}

After having performed the
 integration over $r$:
\bea 
\begin{array}{l}
\qquad{\displaystyle{{d\Delta\over dE}}}={\displaystyle{{d\Delta_{in}\over dE}}}
+{\displaystyle{{d\Delta_{out}\over dE}}}\\
 {\mathrm{where}}  \\
\qquad {\displaystyle{{d\Delta_{in}\over dE}}}={\displaystyle{\int_0^{R}}}\ 
 dr\ 
 {\displaystyle{{d\delta_{in}\over dE}}}
\qquad {\mathrm{and} }\qquad
 {\displaystyle{{d\Delta_{out}\over dE}}}=
 {\displaystyle{\int_R^{r_{max}}} dr\  
 {\displaystyle{{d\delta_{out}\over dE}}}\ ,}
\end{array}
\label{den}
\eea
we compare in fig. \ref{comp2-1} eqs. (\ref{df}) and (\ref{den}). 
We can see from the figure that the correction to our
 naive formula (\ref{naive}) is relatively very small and 
 better for larger energies \footnote{The falling down of the dashed line 
 in fig. \ref{comp2-1} for $E<-8|b|$ is a truncation artefact.}.

\begin{figure}[h]   
\begin{center}\mbox{\epsfig{file=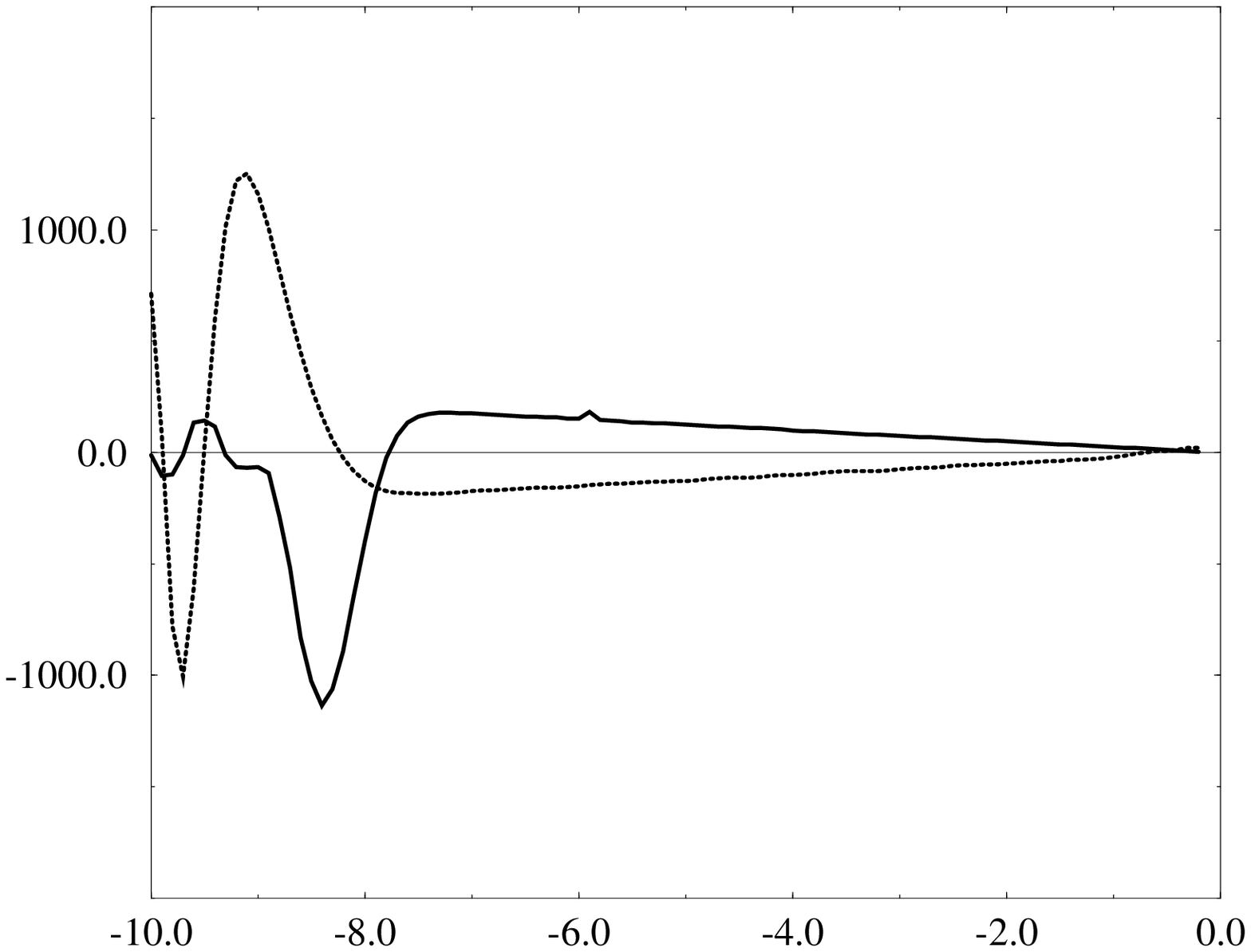,height=8cm}}
\caption[]{\it The horizontal axis represents the energy
   in units of $|b|$. The summation over $l$ has been
performed from $1$ to $80$. For $E > 8b$,
 the positive curve represents (eq. (\ref{den})) the contributions 
  ${d\Delta_{out}} /{dE}$
  of the exterior of the star, the negative curve the contributions 
  $\frac {d\Delta_{in}}{dE}$ of the interior. For the
exterior contribution, $r$ have been integrated from $R$ to $3 R$, the
star radius $R$ being fixed to $R=10 |b^{-1}|$. An accurate cancellation
of both pieces is seen. The oscillation seen for $E < 8b$ is due to the
truncation artefact.
} \protect\label{comp2}
\end{center}
\end{figure}

 The integration over $r$, where $r_{max} > R$, shows that 
\bea
{d \Delta_{in}\over dE} <0\ \ {\mathrm{and}}\ \ 
{d\Delta_{out}\over dE} >0,
\eea
\noindent 
 which means that 
  our crude estimate is an underestimate (overestimate) of the probability 
  density inside (outside) the star. This fact is due to the tendency of
   the exact solution to provide a continuous matching on the border, 
   while our ``crude'' formula has a jump. We shall discuss this in more
    detail.  Moreover, we have  
found that this overestimate and underestimate approximately  cancel
  as 
can be seen from fig. \ref{comp2}. Note that in this figure\\
 (fig. \ref{comp2}), the summation over $l$ goes to $l_2=80$, the
fluctuations that we see for $E<-8|b|$ are due to the  truncation in the
sum  over $l$ when $|E|\ R  > 80$.

Let us look in more detail into how the sudden jump of our naive formula 
(\ref{naive}) 
 when crossing the border of the star is smoothed 
 down into the exact result.  
There must be some layer in which the transition takes place. 
As we have already claimed, this layer is dominated  
by the functions in the domain of eq. (\ref{interm}).   
  The width of the border effect depends directly on 
the speed of the falling-down of the Bessel functions in (\ref{interm}).  
If we use the fact that, for $l$ high enough, the spherical Bessel functions
behave like $(|k|\  r)^l$ (see eq. (\ref{comportement})), we can try  
an estimate of the border
 effect, which gives the correction to our ``naive"  eq. (\ref{naive}).
 The mean value of $l$ inside the domain (\ref{interm}) is
\bea
l_{{\mathrm{mean}}}= R \ {\bigg{|}}\left(E-{\frac b 2}\right){\bigg{|}}\ .
\label{lmoy}\eea
 
 If we suppose that half of the jump occurs at $r=R_-\simlt R$ in the
interior of the star and the other half outside  $r=R_+\simgt R$, we can construct a
term proportional to 
\bea
{dS(r)\over dE}\equiv \ 4\pi r^2   {E^3 \over 2 \pi^2}\left[{\frac 1 2 }(1-\alpha_t^2)
\left ( {r\over R}\right )^{2 l_{{\mathrm{mean}}}}\theta(R-r)-
{\frac 1 2 }(1-\alpha_t^2)
\left ({R\over r}\right )^{2 l_{{\mathrm{mean}}}}\theta(r-R)\right]\ , \label{surf}\eea 
 which will have  the role
of smoothing the jump in our naive approximation formula eq. (\ref{naive}).

\begin{figure}[h]   
\begin{center}\mbox{\epsfig{file=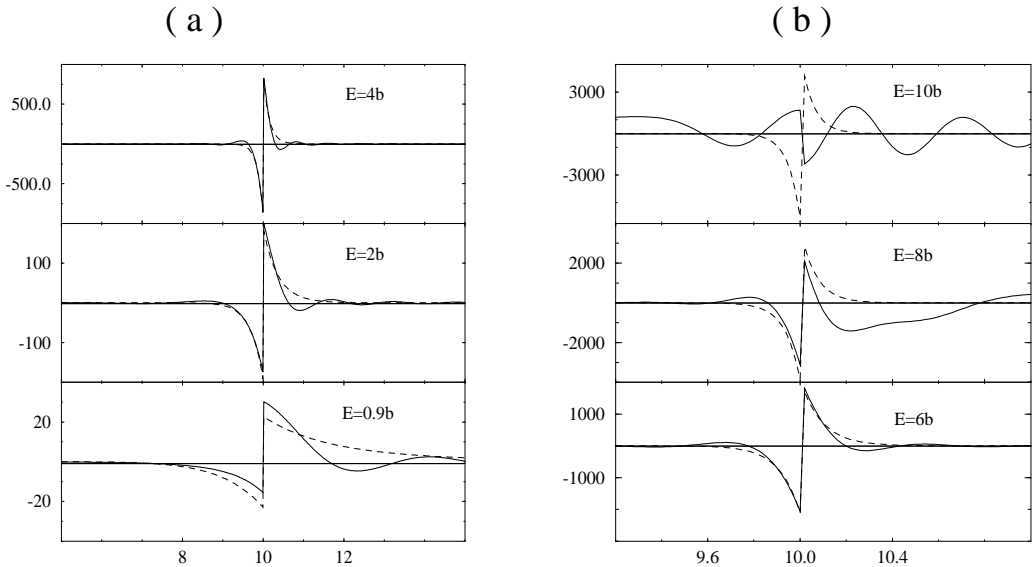,height=8cm}}
\caption[]{\it{The dashed line represents ${\displaystyle{{
dS\over dE}}}$ in 
eq. (\ref{surf}) and the solid line  
${\displaystyle{{d\delta \over dE}}}$ in eq.
(\ref{delta}), 
 as functions of the position $r$, in units of $|b^{-1}|$ 
 (the radius of the star 
has been fixed to $R=10\ |b^{-1}|$) 
for different values of the energy. The summation
over $l$ has been done from $1$ to $80$. The two 
lines are similar, confirming the validity of approximation
(\ref{surf}), except for $E=10 |b|$ where the truncation artefact
dominates. }} \protect\label{comp5}
\end{center}
\end{figure}

In order to confirm this effect, we have compared 
in fig. \ref{comp5}, for different values of the energy, ${dS/ dE}$ with
${d\delta/dE}$.

In fig. \ref{comp5}, 
 we can see the agreement between the l.h.s. of eqs. (\ref{surf}) and  (\ref{delta}), 
  except for $E$ such that $|E|\ R \simgt l_2$, 
where the truncation
effects are important.

To have an estimate of the width $d$ in which the ``joining" occurs, it
suffices to write \bea \left( \frac r R \right )^{l_{{\mathrm{mean}}}} = \exp\left (
l_{{\mathrm{mean}}} \log( \frac r R) \right )\ ; \label{raccord}\nn \eea  
defining $d$ such that the probability decreases by a factor of
$1/e$, we get \bea d= \frac 1 {|E- b/2|} \ .  \eea 
 ~From 
fig. \ref{comp5}, we can easily see that the larger  $|E|$ is, the
narrower  the width of the joining layer is.
 This is clearly seen in fig. \ref{comp5} a while in \ref{comp5}  b,
  beyond $E=8 b$,
the truncation effect comes into play, since $|E| \ R\simgt l_2=80$.

\newpage

\section{discussion}\label{discussion}

 We  now have a qualitative understanding of the dominant contribution to the 
 star
 mass correction due to the neutrino exchange. It doesn't vanish when we
 consider the realistic $(3\ +\ 1)$-dimensional star, because of a border
 effect proportional to the volume of the star. It is ultraviolet- divergent
  and we have to get some understanding of the ultraviolet cut-off
 $C$. 
 This cut-off corresponds to the limit up to which our theory describes, to a
 good approximation, the exact one. The dominant contribution (see eq. (\ref{naive})) 
 $w_{crude}(r)$ to the total weak energy is equal to 
 (see eqs. (\ref{naive}) and (\ref{dfcrude})):
\bea W_{crude}=
{\displaystyle{\frac 2 {3 \pi} R^3 \int_{-C}^0 dE\  E ((E-b)^2 -E^2)}} \simeq 
 -\frac 4 {9\pi} R^3 C^3 b \ .\label{43}
\eea
This result presents the unexpected feature to be odd in $b$, 
while in the perturbative calculation  
the contributions with an odd number of neutrons vanish exactly, 
resulting in a total energy even \footnote{We are indebted to 
Ken Kiers and Michel Tytgat
  for having pointed out this fact to us.} in 
$b$
. 
It is worth noticing that 
 had we started from the symmetrized expression for the vacuum 
 loop eq. (\ref{sommesym}), we would have obtained, instead of (\ref{43}),
\bea W_{sym-crude}=
{\displaystyle{\frac 2 {3 \pi} R^3 \int_{-C}^0 dE\ 2 E b^2}} \simeq 
 \frac {-2} {3\pi} R^3 C^2 b^2 \ .\label{symcru}
\eea

The effective Lagrangian (\ref{lageff})  
is valid only under the assumption that the neutrons are approximately static.
This breaks down, of course, for an energy scale of, say, $C\sim 100$ MeV.
Beyond that scale,  the neutrons feel the recoil of the
scattering. Not far
above, one encounters the scale  of confinement in QCD $\sim 1\ 
{\mathrm {GeV}}$, and the interaction start to ``see'' the substructure of
quarks and gluons. 

 At about the same scale, $C\sim {1\over r_c}$, where $r_c\simeq 0.5\ 
 {\mathrm {fm}}$, as noticed by Fischbach \cite{fischbach}, a repulsive core
  prevents the
 neutron from ``piling-up" in space. In the same ballpark, the
 average distance between neutrons $1/C = n_n^{-1/3}\sim 1 $ fm ($n_n$ is the 
 neutron 
 density in the neutron star, $n_n\simeq 0.4 \ \mathrm{fm}^{-3}$) is the 
 inverse of an ultraviolet
  cut-off since, 
 for smaller distances, our picture of an homogeneous
 background of neutrons breaks down: the neutrino ``see'' individual neutrons
 in a vacuum that is the standard vacuum. 
 All these cut-offs are of the same order of magnitude. Choosing the latter,
  $C= n_n^{1/3}\sim 1 $ fm$^{-1}$, we get from
 (\ref{43}) per unit volume an energy of 
 
 \begin{equation}  W_{crude}\simeq
-\frac {b C^3}{3 \pi^2} \simeq \frac{|b| n_n}{20} \sim 10^{-9}\ 
{\mathrm{GeV}}/{\mathrm{fm}}^3\simeq
 \  10^{-8} m_n\  {\hbox{ per neutron}},\label{estimate}
 \end{equation} 
 $m_n$ being the neutron mass, 
 which means an {\it interaction energy} eight orders of magnitude 
 below the neutron mass, and thus 
 totally negligible. 
 ~From (\ref{symcru}),  the result would be even smaller:
 \begin{equation}  W_{sym-crude}
 \simeq -\frac {b^2 C^2}{2 \pi^2} \simeq -\frac{|b|^2 n_n}{40 C} \sim -10^{-16}\ 
{\mathrm{GeV}}/{\mathrm{fm}}^3\simeq
 \  -10^{-15} m_n\  {\hbox{ per neutron}}.\label{estimate-sym}
 \end{equation} 
 We did not yet manage to understand theoretically 
 the relation between the results (\ref{estimate}) and 
 (\ref{estimate-sym}). They are derived respectively from eqs. (\ref{somme}) 
 and (\ref{sommesym}). On the one hand, eq.  (\ref{sommesym}) leads to a result 
 even in $b$, as does the perturbative
 expansion;  on the other hand, taking the negative energies for the vacuum
 seems more physical to us. It should also be noted that eq. (\ref{sommesym}) 
 is equivalent in Fourier space (see eq. (\ref{fourier})), 
  to averaging the complex 
 integration closed in both the upper  and in the lower half-planes. 
  The issue is clearly
 related to the ultraviolet regularization. Indeed, subtracting
 (\ref{sommesym}) from (\ref{somme}) and using the closure theorem, it is
 easy to obtain the formal result: 
 \bea w(x)-w_{sym}(x)= \frac i 2 \frac
 d {dx_0} \delta_3(\vec x -\vec y) =0 \label{zer0} \ , \eea 
 where the
 vanishing occurs since we derive on time a time-independent quantity.
 Once regularized by an ultraviolet cut-off so that $-C < E < C$ as in
 (\ref{43}) and (\ref{symcru}), $w(x)-w_{sym}(x)$ is not zero. We therefore
 conclude that the difference between (\ref{estimate}) and
 (\ref{estimate-sym}) is related to different ultraviolet
 regularizations. 
 At this point we prefer to present both results, while studying  
 the issue further.

 Another argument leads to reduce the estimate (\ref{estimate}). Indeed, 
 throughout this
 paper we have considered stationary eigenstates of the Hamiltonian that  
  extend over all space, such as plane waves.  The use of 
 such extended waves 
 contains the implicit assumption that the neutrino wave packets extend 
over the whole star and far beyond. In other words our solutions know about 
the
whole star and its surroundings. This is perfectly legitimate for low-energy
neutrinos, since their mean free path is much larger than the radius of the
star. This means that they feel the coherent interactions from the neutrons as
expressed by the effective Lagrangian (\ref{lageff}), 
but they do not experience
incoherent scattering with the neutrons of the star. 

 A rough estimate of the neutrino cross section with neutrons is
 $\sigma \sim E_\nu^2 G_F^2$, where $G_F$ is the Fermi constant and $E_\nu$ the
 neutrino energy. The mean free path is approximately
 
 \begin{equation}
 \lambda \simeq \frac 1 {n_n \sigma}\sim \frac 1 {n_n E_\nu^2 G_F^2}\simeq
 \left( \frac {{0.5 \ \mathrm {MeV}}}{E_\nu}\right )^2 {\mathrm{km}} 
 \simeq \left( \frac {{100\  \mathrm {keV}}}{E_\nu}\right )^2 25\ {\mathrm{km}}
  \ ,\label{mean}
 \end{equation}
 where $n_n$ is the neutron density in the star. This means that for an energy
 of 100 keV, the mean free path is $\sim 25 $ km, of the order of the star
 radius. Let us call this auxiliary ultraviolet cut-off $c$ ($c\simeq 100$ keV).
   
 We now decompose the integral on the negative energy modes in two parts. 
 For $|E_\nu|<c$ we use formula (\ref{43}). For energies larger than $c$ but 
 smaller than $\sim 0.2$ GeV, we may argue that the wave packets must have a size of the order of the mean free path. 
 Therefore the ``knowledge'' of the border effects extends
 inside the star on a distance of the order of $\lambda$ from the border.
 A neutrino deeper in the star does not feel the
 border, we are back to the situation of an infinite star \cite{rescue} where we had a vanishing result. The
 non-zero contribution comes only from a thin region around the border of width 
 $\lambda$.
 Instead of $R^3$ in (\ref{43}), we take $R^2 \lambda \simeq (c/E_\nu)^2 R^3$ 
 (using the $1/E_\nu^2$ law in (\ref{mean}) and the fact that for
 $c=|E_\nu|$, we have  
 $\lambda \simeq R$). We thus have,  instead of (\ref{43}):
 \begin{equation}
 \frac 2 {3\pi} R^3 \int_{-c}^0 dE\  (-2) E^2 b + \frac 2 \pi  R^3 
 \int_{-C}^{-c} \ dE  (-2 c^2) b = \frac 4 \pi R^3 b (c^3/3+c^2(C-c))\simeq 
 R^3 c^2 C b \ .    
 \end{equation}

 Since $c/C \sim 10^{-3}$, we get an average estimate for the energy per volume 
 still six  orders of magnitude below the result of (\ref{estimate}), i.e.

\begin{equation}
 \sim  10^{-14} m_n \hbox{ per neutron}.
 \end{equation}

 Finally it should be remarked that all along this work we have solved a
 stationary problem. This means that we have assumed the star, and 
 also the neutrino states inside it,
  to have reached an equilibrium status. It is well-known that the stars evolve
  during their life; we thus have implicitly assumed an adiabatic adjustment of
  the neutrino states. Since the star evolution is slow and since neutrino
  motion inside them has the velocity of light, this seems to us a reasonable
  assumption. Some further study might still be welcome. We have
also assumed
  a zero temperature for the neutrinos, since we believe that their
interaction is too small for them to thermalize.

\section{Concluding remarks}\label{remarks}
In order to settle definitively the question of the stability of a neutron star,
 the multibody exchange of massless 
 neutrinos has been computed analytically and numerically for a finite star. 

 The effect of a border is twofold. First it induces in a natural way the
 neutrino condensate as proved in  \cite{note}. 
The latter condensate, does not produce any neutrino exchange interaction energy
in the simplified $(1\  +\  1)$-dimensional case and we find it to be negligible in the
realistic $(3\  +\  1)$-dimensional case. 

The second effect of the border is that the neutrino zero point energy inside
the star differs from the outer one because of  negative-energy waves  that cannot
penetrate inside the star, being beyond the limiting refraction index at the
border. This
 contribution  is proportional to the volume of the star, but it is still  tiny ($10^{-8}$--$10^{-13}
 \ {\mathrm{GeV}\hbox{ per neutron}}$), completely negligible in 
 comparison with the neutron mass.
 {\it We find no infrared divergences in the full non-perturbative result,
  which would have
  necessitated the
 introduction of a neutrino mass.}

The general conclusion of this work is that the neutrino does not need to be
massive to ensure the stability of a neutron star. This is in agreement with 
recent works (commented below) by Kachelriess \cite{Kachelriess} and by Kiers \& Tytgat
\cite{kiers}. 
There is no
catastrophic effect due to the multibody massless neutrino exchange. 
As already stated in refs. \cite{rescue} and \cite{note}, this catastrophic result claimed in
\cite{fischbach} is
only due to an attempt to sum up the perturbative series  outside its radius of
convergence.

While finishing this paper, there appeared a paper by Kiers and  Tytgat 
\cite{kiers}. They accept the point of view developed in \cite{rescue} for an 
infinite star and try, as we do, to solve the problem of a finite star. 
 Their starting goal is, as ours, to compute the density in eq. (\ref{somme}).
They use a clever technique based on quantizing in a large sphere and 
expressing the vacuum energy density in terms of the phase shifts. They 
first study
analytically the unphysical but illustrative  case of small $bR$ and 
then numerically the large $bR$ case. They show that the perturbative
 series 
{\it \`a la Fischbach} already breaks down as early as 
 $b R> \pi$ (for the neutron star, $b R\sim 10^{12}$)
   while {\it the non-perturbative calculation gives a negligible result,
    a conclusion which we fully share}. One difference between their 
    result and ours is that they find a relation between the energy 
    density of the neutron star and that of the neutrino condensate. 
    We find on the contrary that the result is mainly due to non-penetrating 
    negative-energy waves, which are not related to the condensate. 
   We did not yet manage to understand the reason for this discrepancy. 
   It might be related to different UV-regularization methods.

  Kachelriess \cite{Kachelriess} also agrees with us
about Fischbach's ``catastrophic'' result. He
computed the total weak self-energy for an infinite neutron star 
following the Schwinger method, by using the neutrino propagator in
momentum space, as we did in ref. \cite{rescue}. He
obtained a non-zero weak self-energy without taking into account the
neutrino sea effects. As we acknowledged in section \ref{flat} and in
ref. \cite{note},  a minor mistake was made in
ref. \cite{rescue}: the contribution of a pole had been forgotten
in the calculation. Once this mistake is corrected,
we agree  with Kachelriess about the result when neglecting the neutrino sea.
 He attributed the discrepancy to the fact that we took the limit
($y \to x$) before integrating over the whole space to obtain the total
weak self-energy (see eq. (\ref{Wanalyt})). We do not agree with this
 conjecture
about the discrepancy, first because, as we just mentioned, the
 forgotten pole removes the discrepancy, second because we 
have verified that our previous UV-regularization makes
  the $x\to y$ limit regular.
 Finally, from our analysis of finite stars \cite{note},
  we insist that the condensate {\it has to be taken into account}
   and, amazingly, it exactly cancels the forgotten pole contribution, 
   resulting finally in $w(\vec x)=0$ for an infinite star.

   A few weeks later, Fischbach and Woodahl \cite{repeated} repeated 
   Fischbach's original claim and
   used the same expansion, order by order, in the number of neutrons,
i.e. equivalently in perturbation in the parameter $bR$.
    Astonishingly enough they did not consider the series of works
demonstrating that this series is simply divergent, but that the total
result may be computed directly by the effective Lagrangian technique.
They argued that our non-perturbative calculation encounters
    cancellations because, in the effective approach, the neutron medium
is assumed by us to be a continuous background. Of course, treating the
neutron medium as a homogeneous continuum medium is an approximation \`a
la Hartree--Fock, and it should be corrected {\it in the ultraviolet} by
taking into account the
    correlation between neutrons. This is precisely one of the reasons
why  we considered
    that a natural ultraviolet cut-off was the energy scale of a few
MeV. The authors of \cite{fischbach} take the
size of the neutron hard core as an ultraviolet cut-off. Why not,
although many other ultraviolet effects arise at the same scale of a
few 100 MeV: the neutron recoil, the
    quark and gluon content of the neutron, without forgetting the
incoherent neutrino--neutron scattering discussed in the previous 
section.
     But these ultraviolet effects {\it will not at all modify the
analysis of the infrared catastrophe} advocated by Fischbach and denied
by us.
 The authors of \cite{repeated} seem to imply that we have added some
unjustified assumption in our work. The truth is on the contrary that we
have
 assumed nothing that they did not assume themselves, such as the static
neutron assumption, but we have not assumed, as they do, that a neutron is
not allowed to interact more than once, neither did we make the drastic
approximations that appear in their work at high order in
perturbation. {\it The effective Lagrangian approach allows to compute
exactly, in a simple manner, and with fewer assumptions than the
perturbative expansion approach.}

 The authors of \cite{repeated} criticize our recent $(1\  +\  1)$-dimensional
  toy calculation \cite{note} arguing that the critical
parameter $bR$ is much smaller than 1 in $(1\  +\  1)$
dimensions.  However, they did not notice that
 our $(1\  +\  1)$-dimensional result is absolutely exact, independently
of the parameter $bR$, which, incidentally, we have taken to be large.

    To finish, we feel it necessary to insist. The main issue is the
failure of the perturbative expansion, which is infrared-divergent. 
Happily one can spare
    this difficulty thanks to the effective action technique. Once this
point is understood, the different analyses all agree, notwithstanding
minor discrepancies, that
    {\it although the
    massless neutrino exchange between fermions is a long-range
interaction, it
    does not give any significant contribution to the total energy of a
neutron
    star, finite or infinite}.

\section*{Acknowledgements}

We are specially indebted to  M. B. Gavela for the 
helpful discussions that initiated the work and for
reading the manuscript. We wish to thank K. Kiers and M. H. G. Tytgat
 for reading the manuscript and very important comments on our draft. We thank
 also M.  Kachelriess for his interest in our work and his helpful comments.  
J. Rodr\'{\i}guez--Quintero thanks M. Lozano for his invaluable support.
 This work has been partially supported by Spanish CICYT, project 
PB 95-0533-A.


\begin{thebibliography}{99}
\bibitem{Fein68} G. Feinberg and J. Sucher, Phys. Rev. {\bf 166} (1968) 1638.
\bibitem{Fein89} G. Feinberg, J. Sucher and C. K. Au, Phys. Rep. {\bf 180} (1989) 1.
\bibitem{sikivi} S. D. H. Hsu and P. Sikivie, Phys. Rev. {\bf D49} (1994) 4951.
\bibitem{rescue} {As.Abada, M.B. Gavela and O. P\`ene, Phys. Lett. {\bf B387}
(1996) 315.} 
\bibitem{Kachelriess} M. Kachelriess, {\it Neutrino self-energy and pair
creation in neutron stars}, hep-ph/9712363, to be published in Phys.Lett.B.  
\bibitem{kiers} K. Kiers and M. H. G. Tytgat, {\it The neutrino ground state in a
macroscopic electroweak potential}, hep-ph/9712463.  
\bibitem{fischbach} {H. Kloor, E. Fischbach, C. Talmadge and G. L. Greene, 
Phys. Rev. {\bf D49} (1994) 2098, E. Fischbach, Ann. Phys. {\bf 247} (1996)
 213,\\
B. Woodahl, M. Parry, S-J. Tu and E. Fischbach, hep-ph/9709334, hep-ph/9606250.}
\bibitem{smirnov} A. Y. Smirnov and F. Vissani, hep-ph/9604443, {\it A
Lower Bound on Neutrino Mass}, Moriond Proceedings (1996). 
\bibitem{loeb} A. Loeb, Phys. Rev. Lett. {\bf 64} (1990) 115.

\bibitem{note} As. Abada, O. P\`ene and J.
Rodr\'{\i}guez--Quintero, {\it Multibody neutrino exchange in a neutron star:
neutrino sea and border effects}, CERN-TH/97-350, FAMNSE-97/19, LPTHE-Orsay-97/65, 
hep-ph/9712266, to be published in Phys.Lett.B.
\bibitem{smirnov1} A. Smirnov,  private communication.
\bibitem{wolf} L. Wolfenstein, Phys. Rev. {\bf D17} (1978) 2369; P. Langacker,
J.P. L\'eveill\'e and J. Sheiman, Phys. Rev. {\bf D27} (1983) 1228.
\bibitem{nieves} J.C. D'Olivo, J.F. Nieves and M. Torres, Phys. Rev. {\bf
D} (1992) 1172; C. Quimbay and S. Vargas-Castrill\'on, Nucl. Phys. {\bf B451}
(1995) 265.
\bibitem{schwinger} J. Schwinger, Phys. Rev. {\bf 94} (1954) 1362.
\bibitem{Mor97} J. Rodr\'{\i}guez--Quintero, {\it Resurrection
of a star}, to appear in Moriond Proceedings (1997).
\bibitem{Gav94} M. B. Gavela, M. Lozano, J. Orloff and  O. P\`ene, Nucl.
Phys. {\bf B430} (1994) 345.
\bibitem{Rod97} J. Rodr\'{\i}guez--Quintero, O. P\`{e}ne and M. Lozano,
Ann. Phys. (N. Y.) {\bf 259} (1997) 65-96. 

\bibitem{watson} G. N. Watson, {\it Theory of Bessel Functions} (Cambridge
University Press, Cambridge, 1966).
\bibitem{greiner} W. Greiner, 
{\it Relativistic quantum mechanics: wave equations} (Springer Verlag,
Berlin-
Heidelberg, 1990). 

\bibitem{repeated} E. Fischbach and B. Woodahl, {\it Neutrino-exchange interactions in 
1, 2
and 3 dimensions}, hep-ph/9801387.  





\end{thebibliography}
\end{document}